\documentclass[review]{elsarticle}

\usepackage{lineno,hyperref}
\usepackage{hyperref}
\usepackage{fancyhdr}
\usepackage{amsmath}
\usepackage{mathtools}
\usepackage{amsfonts}
\usepackage{amssymb}
\usepackage{textcomp}
\usepackage{booktabs}                      % tables
\usepackage{multicol}
\usepackage{multirow}
\usepackage{nomencl}
\usepackage{caption}
\usepackage{float}
\usepackage{rotating}
\makenomenclature
\usepackage{tabularx}                      % tables with predefined width
\usepackage{natbib}
\usepackage{adjustbox}
\usepackage[graphicx]{realboxes}
\usepackage[toc,page]{appendix}
\usepackage{cancel}
\usepackage{graphicx}
\usepackage{subcaption}
\usepackage{epstopdf}
\usepackage{tikz}
\usetikzlibrary{shapes,arrows}
\modulolinenumbers[5]
\usepackage{color}
\usepackage[normalem]{ulem}
\newcommand{\ns}[1]{{#1}}
\newcommand{\pc}[1]{{#1}}
\newcommand{\nf}[1]{{#1}}
\begin{document}

\begin{frontmatter}

%textwidth in cm: \printinunitsof{cm}\prntlen{\textwidth}

%\title{An algebraic volume of fluid method for interface-resolved simulation of evaporating flows}
\title{A volume-of-fluid method for interface-resolved simulations of phase-changing two-fluid flows}

\author[kth]{Nicol\`o Scapin}
\ead{nicolos@mech.kth.se}

%\author[hi]{Pedro Costa\corref{mycorrespondingauthor}}
    \author[kth]{Pedro Costa\corref{mycorrespondingauthor}\footnote{Present address of Pedro Costa: Faculty of Industrial Engineering, Mechanical Engineering and Computer Science, University of Iceland, Hjardarhagi 2-6, 107 Reykjavik, Iceland}}
\ead{p.simoes.costa@gmail.com}
\cortext[mycorrespondingauthor]{Corresponding author}

\author[kth,ntnu]{Luca Brandt}
\ead{luca@mech.kth.se}

\address[kth]{Linn\'e FLOW Centre and SeRC (Swedish e-Science Research Centre), KTH  Mechanics, SE-100 44 Stockholm, Sweden}
%\address[hi]{Faculty of Industrial Eng., Mechanical Eng. and Computer Science, University of Iceland (HI), Reykjavík, Iceland}
\address[ntnu]{Department of Energy and Process Engineering, Norwegian University of Science and Technology (NTNU), Trondheim, Norway}

\begin{abstract}
We present a numerical method for interface-resolved simulations of evaporating two-fluid flows based on the volume-of-fluid (VoF) method. The method has been implemented in an efficient FFT-based two-fluid Navier-Stokes solver, using an algebraic VoF method for the interface representation, and extended with the transport equations of thermal energy and vaporized liquid mass for the single-component evaporating liquid in an inert gas. The conservation of vaporizing liquid and computation of the interfacial mass flux are performed with the aid of a reconstructed signed-distance field, which enables the use of well-established methods for phase change solvers based on level-set methods. The interface velocity is computed with a novel approach that ensures accurate mass conservation, by constructing a divergence-free extension of the liquid velocity field onto the entire domain. The resulting approach does not depend on the type of interface reconstruction (i.e.\ can be employed in both algebraic and geometrical VoF methods). We extensively verified and validated the overall method against several benchmark cases, and demonstrated its excellent mass conservation and good overall performance for simulating evaporating two-fluid flows in two and three dimensions. 
\end{abstract}


\end{frontmatter}

\linenumbers

\section{Introduction}
Multiphase flows undergoing phase change are found in many environmental and industrial contexts, such as cloud formation and rainfall, cooling towers, wet scrubbers and spray combustion. These systems are inherently complex, with the different phases exchanging mass, momentum and energy through an interface that moves and deforms with the flow. The thickness of this interface is typically several orders of magnitude smaller than any other relevant flow scale. From a continuum mechanics modeling perspective, the interface is often modeled as infinitesimally thin, and its physics simplified to appropriate interphase coupling conditions, derived from interfacial transport balances and thermodynamic considerations \cite{ishii2010thermo}. Even considering these modeling assumptions, first-principles, interface-resolved numerical simulations of these flows remain a challenge although simulations can provide valuable insights for e.g.\ upscale models aiming to improve predictive engineering tools \cite{kharangate2017review} in a context where experiments are particularly difficult.\par
Interface-resolved simulations of two-fluid flows are often divided into two categories according to the interface representation \cite{elghobashi2019direct}: (1) \emph{Interface-tracking} (or \emph{front-tracking}) methods explicitly define a mesh with Lagrangian markers, attached to and moving with the interface; (2) \emph{interface-capturing} methods define the interface in a higher dimension by solving a transport equation for an auxiliary scalar field. Front-tracking methods \cite{unverdi1992front} allow in general for a more accurate interface representation, at the cost of more complex implementations, specially when it comes to handling surface topology changes. \ns{For the same grid resolution and computational cost, interface-capturing methods are in general less accurate in the calculation of the interface properties, such as curvature and normal vector, but simpler to implement, and naturally handle topology changes thanks to the implicit interface representation}. Methods that fall in this category are the level-set, \cite{osher2006level}, the volume-of-fluid \cite{tryggvason2011direct}, constrained-interpolation \cite{yabe2001constrained} and diffuse-interface \cite{anderson1998diffuse} methods. Given the pros and cons of the different approaches, most of them have been used to simulate phase-changing two-fluid flows in different configurations. Also, most methods assume that one of these two dominant mechanisms drives phase change: (1) large temperatures, when phase change is triggered by a prescribed interface saturation temperature -- \emph{boiling}, and (2) species concentration gradients near the interface, with phase change induced by a prescribed non-uniform interface concentration -- \emph{evaporation}.\par
Front-tracking (FT) methods have been used to study boiling flows, with application to film boiling; see e.g.\ \cite{juric1998computations},~\cite{esmaeeli2004computations_i} and~\cite{esmaeeli2004computations_ii}. Recent studies have extended this framework to evaporating two-fluid flows in two dimensions \cite{irfan2017front}, also in presence of chemical reactions \cite{irfan2018front}. Despite the successes of FT methods for phase change, highly scalable parallel implementations are challenging and remain scarce \cite{farooqi2019communication}. Such a feature is crucial for simulating e.g.\ turbulent gas-liquid flows, which may require massive simulations with $\mathcal{O}(10^8-10^9)$ Eulerian grid cells \cite{ling2019two}.\par
\pc{In regard to} interface-capturing methods, the first study of interface-resolved simulations of phase changing two-fluid flows used a level-set method, applied to film boiling \cite{son1998numerical}. Several studies have followed, aiming to incorporate interphase-coupling jump conditions with the so-called ghost-fluid method, which provides a sharp representation of the jump at the discrete level. These methods have been employed for boiling \cite{gibou2007level,tanguy2014benchmarks,lee2017direct}, evaporation \cite{tanguy2007level} and the combination of the two \ns{\cite{villegas2016ghost}}. Despite the proven successes of the level-set method for phase change problems, these are not mass-preserving by construction. \ns{Mass conservation properties} are desirable for numerical simulations of several systems, e.g.\ in multiphase turbulent flows, where the flow statistics should be collected over periods of time long enough that mass loss may become significant. Recent studies have dealt with this problem of level-set methods (in the absence of phase change) by introducing mass correction steps, where lost mass is redistributed near the interface depending on e.g.\ the local interface curvature \cite{luo2015mass,ge2018efficient}.\par
Another widely-used interface-capturing approach for multiphase flow simulations is the volume-of-fluid (VoF) method, which has the major strength of ensuring mass conservation by construction. These methods have been extended to simulations of boiling flows \cite{welch2000volume,agarwal2004planar,yuan2008numerical} (in two dimensions), and evaporation \cite{schlottke2008direct,hardt2008evaporation}. Mass-preserving methods for boiling flows have also been devised based on coupled level-set and VoF methods \cite{tomar2005numerical}, or the constrained-interpolation method \cite{sato2013sharp}. Most of these methods use the flow velocity to transport the liquid/gas volume fraction field and include a source term accounting for the phase change. This can however cause numerical issues, as the velocity in the presence of phase change has a jump across the interface, and its divergence is non-zero. Moreover, these approaches cannot be easily and directly adapted to VoF methods where a smooth (and often divergence-free) velocity field is necessary to transport the VoF function with preserved interface thickness. Interface smearing due to the transport of a VoF function with a non-divergence-free velocity field is often avoided by resorting to ad-hoc interface-compression schemes developed for certain classes of algebraic VoF methods \cite{ubbink1999method,hoang2013benchmark}. Other approaches using VoF for evaporating flows rely on the geometrical reconstruction of the interface for computing interface mass fluxes and estimating and re-distributing divergence errors in the grid cells around the interface in order to improve mass conservation; see \cite{schlottke2008direct}. Finally, we should note that, in the context of numerical studies for evaporating flows, thorough verification/validation studies demonstrating the grid convergence of the methods for different benchmarks remain scarce.\par
Here, we present a numerical model for three-dimensional direct numerical simulations of evaporating flows using a volume-of-fluid method. 
The transport of vaporized liquid mass is solved with a Dirichlet boundary condition at the interface, obtained assuming thermodynamic equilibrium through the Clausius-Clapeyron relation, and solved with the aid of a reconstructed level-set field. This allows to easily compute the interfacial mass flux in a band around the interface \cite{tanguy2007level}. 
\ns{We propose here to transport the VoF function with a smooth interface velocity consisting of two terms: a divergence-free extension of the liquid velocity field, and an irrotational term due to phase change. Accordingly, the standard directional-splitting method used for the VoF advection is extended with a volume deflation step. This results in a novel approach for transporting the VoF function that shows excellent mass conservation properties, and can be easily applied to other geometrical or algebraic VoF methods for incompressible two-fluid flows. Unlike previous approaches, this allows to use a whole-domain formulation for the momentum equations, \pc{without introducing interface-sharpening terms in the VoF transport equation}.}
The method is implemented in an efficient, FFT-based two-fluid finite-difference Navier-Stokes solver, extended with the MTHINC (algebraic) VoF method for the interface representation.
We verify the proposed method against several benchmark cases of droplet evaporation, and demonstrate the solution grid convergence. Moreover, we validate the overall numerical method against psychrometric data, and prove its ability to simulate evaporating flows in the presence of large droplet deformations near solid boundaries, in two and three dimensions. \par
This paper is organised as follows. The governing equations are presented within the so-called one-fluid formulation in section~\ref{sec:gov_eqn}. Then section~\ref{sec:num_meth}, describes the numerical approach used to solve the system of equations, together with the interface representation and construction of the interface velocity. The overall method is verified and validated against several benchmarks in section~\ref{sec:results}. Finally, conclusions are drawn in section~\ref{sec:concl}.%, with potential applications of the proposed methodology.
\section{Governing equations}\label{sec:gov_eqn}
We shall consider a system with two immiscible and incompressible Newtonian fluids: a single component liquid (phase $1$) and an ideal mixture of an inert gas and vaporized liquid (phase $2$). The two phases are bounded by an infinitesimally small interface, through which energy, momentum and mass can be transferred. Evaporation (i.e.\ mass transfer due to phase change) can occur, and is driven by the partial pressure of the inert gas in phase $2$.\par
Before introducing the governing equations, it is convenient to define a phase indicator function $H$ distinguishing the two phases at position $\mathbf{x}$ and time $t$:
\begin{equation}
	H(\mathbf{x},t) = \begin{cases}
						1 \hspace{0.5 cm} \text{if $\mathbf{x} \in \Omega_1$}\mathrm{,} \\
						0 \hspace{0.5 cm} \text{if $\mathbf{x} \in \Omega_2$}\mathrm{,} 
					  \end{cases}
	\label{ind_fun}
\end{equation}
where $\Omega_1$ and $\Omega_2$ are the domains pertaining to phases $1$ and $2$.  
We can use $H$ to define the thermophysical properties in the whole domain $\Omega = \Omega_1 \cup \Omega_2$ as follows:
\begin{equation}
    \xi(\mathbf{x},t) = \xi _1H(\mathbf{x},t)+\xi _2(1-H(\mathbf{x},t))\mathrm{,}
	\label{material_prop}
\end{equation}
where $\xi_{i}$ ($i=1,\,2$ for phases $1$ and $2$) can be the mass density $\rho_i$, the dynamic viscosity $\mu_i$, the thermal conductivity $k_i$ or \nf{the inverse of the heat capacity at constant pressure $(\rho_i c_{p,i})^{-1}$. The reasons behind the use of the harmonic mean for $\rho c_p$ instead of the arithmetic one, as for the other thermophysical properties, will be detailed in section~\ref{num:en_eqn}}. Hereafter, unless otherwise stated, thermophysical quantities not specifically referring to one of the phases are defined from eq.~\eqref{material_prop}.\par
The equations governing the mass, energy and momentum transport for phase $1$ and $2$ are coupled through appropriate interfacial conditions~\cite{ishii2010thermo}. Below we present the governing equations in the so-called one-fluid or whole-domain formulation, where each transport equation is defined in $\Omega$ \cite{prosperetti2009computational,tanguy2014benchmarks}.
\subsection*{Navier-Stokes equations}\label{sec:nseqn}
The interfacial mass flux $\dot{m}$ due to phase change makes the velocity field $\mathbf{u}$ discontinuous. This can be expressed by the following Rankine-Hugoniot condition~\cite{tryggvason2011direct}:
\begin{equation}
	\rho_1(\mathbf{u}_1-\mathbf{u}_{\Gamma})\cdot\mathbf{n}=\rho_2(\mathbf{u}_2-\mathbf{u}_{\Gamma})\cdot\mathbf{n}=\dot{m}\mathrm{,}
	\label{rankine}
\end{equation}
where $\mathbf{n}$ is interface normal vector (pointing to $\Omega_2$), $\mathbf{u}_{i=1,2}$ is the fluid velocity in each subdomain, and $\mathbf{u}_{\Gamma}$ the interface velocity ($\Gamma = \Omega_1\cap\Omega_2$). The continuity equation accounting for this condition reads:
\begin{equation}
    \nabla\cdot\mathbf{u}=\dot{m}\left(\dfrac{1}{\rho_2}-\dfrac{1}{\rho_1}\right)\delta_{\Gamma}\mathrm{,}
	\label{mass}
\end{equation}
where $\delta_{\Gamma}\equiv\delta(\mathbf{x}-\mathbf{x}_\Gamma)$ is a three-dimensional Dirac delta function, non-zero at the interface position $\mathbf{x}_\Gamma$.\par
The momentum equation can be written as follows~\cite{tryggvason2011direct}:
\begin{align}
	\rho\left(\dfrac{\partial\mathbf{u}}{\partial t}+\mathbf{u}\cdot\nabla\mathbf{u}\right)=-\nabla p+\nabla\cdot\left(\mu\left(\nabla\mathbf{u}+\nabla\mathbf{u}^T\right)\right)+\rho\mathbf{g}+\sigma\kappa\delta_{\Gamma}\mathbf{n}\mathrm{,}
	\label{mom}
\end{align}
where $p$ is the pressure field and $\mathbf{g}$ the gravitational acceleration; the right-most term accounts for the jump in stress due to surface tension, with $\sigma$ being the surface tension coefficient and $\kappa$ local interface curvature. \pc{Note that equation~\eqref{mom} is coupled to the equation for mass conservation, eq.~\eqref{mass}, and therefore the velocity field must satisfy the constrain on the velocity divergence, which is directly related to the mass flux at the interface.}

\subsection*{Vaporized liquid mass transport}\label{sec:vap_tra}
The transport of vapor mass is only defined in $\Omega_2$, and driven by a standard convection-diffusion equation:
\begin{equation}
	\dfrac{\partial Y_2^l}{\partial t}+\mathbf{u}\cdot\nabla Y_2^l=D_{lg}\nabla^2 Y_2^l\mathrm{,}
	\label{mass_fraction}
\end{equation}
where $Y_2^l$ denotes the mass fraction of vapor in the ideal mixture (i.e.\ vaporized liquid in $\Omega_2$) and $D_{lg}$ is the diffusion coefficient of vapor in the gas. Note that, since we consider a single component liquid, the analogous equation for the liquid phase is trivial, i.e.\ $Y_1^l = 1$ in $\Omega_1$. The interface boundary condition for $Y_{2,\Gamma}^l$ related to the saturation vapor pressure $p_{2,\Gamma}^{l,sat}$ is as follows:
\begin{equation}
    Y_{2,\Gamma}^{l}=\dfrac{p_{2,\Gamma}^{l,sat}M_l}{(p_{t}-p_{2,\Gamma}^{l,sat})M_g+p_{2,\Gamma}^{l,sat}M_l}\mathrm{,}
	\label{eqn:bc_y}
\end{equation}
where $p_{t}$ is the total pressure of the mixture, and $M_l$ and $M_g$ denote the molar \ns{mass} of the liquid and inert gas. By assuming the thermodynamic equilibrium at the interface, it is possible to relate $p_{2,\Gamma}^{l,sat}$ to the local interface temperature $T_{\Gamma}$, through the Clausius-Clapeyron relation~\cite{bergman2016fundamentals}:
\begin{equation}
	p_{2,\Gamma}^{l,sat}=p_{t}\exp\left[-\dfrac{h_{lv}M_l}{R}\left(\dfrac{1}{T_{\Gamma}}-\dfrac{1}{T^{sat}}\right)\right]\mathrm{,}
	\label{eqn:eq_state}
\end{equation}
where $T^{sat}$ is the liquid saturation temperature at ambient pressure $p_{t}$, $R$ is the universal molar gas constant, and $h_{lv}$ the latent heat of phase change. Finally, since in the current work we limit ourselves to a single-component liquid and neglect the gas dissolution in the liquid phase, the mass balance across the interface results in the following condition for $\dot{m}$ \cite{bergman2016fundamentals}:
\begin{equation}
    \dot{m}(1-Y_{2,\Gamma}^l) = - \rho_2D_{lg}\nabla_\Gamma Y_2^l\cdot\mathbf{n}\mathrm{,}
	\label{mass_flux}
\end{equation}
where $\nabla_{\Gamma}$ denotes the gradient at $\mathbf{x}=\mathbf{x}_{\Gamma}$.
\subsection*{Energy transport}\label{sec:en_eqn}
The conservation of thermal energy can be written in the one-fluid formulation as follows:
\begin{equation}
	\rho c_p\left(\dfrac{\partial T}{\partial t}+\mathbf{u}\cdot\nabla T\right) = \nabla\cdot(k\nabla T) 
    -\dot{m}\left[h_{lv}+(c_{p,1}-c_{p,2})(T^{sat}-T_{\Gamma})\right]\delta_{\Gamma},
	\label{energy}
\end{equation}
where viscous dissipation has been neglected. Here $T$ is the temperature field, $c_p$ the specific heat at constant pressure and $k$ the thermal conductivity. The last term in eq.~\eqref{energy} quantifies the jump in enthalpy due to phase change, mostly due to the latent heat $h_{lg}$, but also due to the differences in specific heat between the two phases. Note that the jump in heat flux at the interface can be easily derived by integrating eq.~(\ref{energy}) across $\Gamma$. 

\subsection*{Governing parameters}\label{sec:physical_param}
Eqs.~\eqref{mass},~\eqref{mom},~\eqref{mass_fraction},~\eqref{eqn:eq_state},~\eqref{eqn:bc_y} and~\eqref{energy} can be written in non-dimensional form by introducing proper scaling parameters, namely a reference velocity, length and temperature scales, $u_{ref}$, $l_{ref}$ and $T_{ref}$, together with reference termophysical properties, here taken as those of the gas phase $2$. The non-dimensional governing parameters read:
\begin{equation*}
    \mathrm{Re} =\dfrac{\rho_2u_{ref}l_{ref}}{\mu_2}\mathrm{,} \hspace{0.5 cm} \mathrm{We}=\dfrac{\rho_2u_{ref}^2l_{ref}}{\sigma}\mathrm{,} \hspace{0.5 cm} \mathrm{Fr}=\dfrac{u_{ref}^2}{l_{ref}|\mathbf{g}|}\mathrm{,} \hspace{0.5 cm}\mathrm{Pr}=\dfrac{\mu_{2}c_{p,2}}{k_{2}}\mathrm{,} 
\end{equation*}
\begin{equation*}
    \mathrm{Sc}=\dfrac{\mu_{2}}{D_{lg}\rho_{2}}\mathrm{,} \hspace{0.5 cm} \mathrm{Ste}=\dfrac{c_{p,2}T_{ref}}{h_{lv}}\mathrm{,} \hspace{0.5 cm} \mathrm{Ste}^m=\dfrac{RT^{sat}}{M_lh_{lv}}\mathrm{,} \hspace{0.5 cm} \lambda_M=\dfrac{M_l}{M_g}\mathrm{,}
\end{equation*}
where $\mathrm{Re}$, $We$, $\mathrm{Fr}$, $\mathrm{Pr}$, $\mathrm{Sc}$, $\mathrm{Ste}$, $\mathrm{Ste}^m$ and $\lambda_M$ are the Reynolds, Weber, Froude, Prandtl, Schmidt, Stefan number, a modified Stefan number and the molar mass ratio; in addition to these, one needs to consider  the ratios of the different thermophysical properties, $\lambda_\xi\equiv\xi_1/\xi_2$, where $\xi$ can be $\rho$, $\mu$, $c_p$ or $k$.
\section{Numerical method}\label{sec:num_meth}
The governing equations are solved on a fixed regular Cartesian grid (i.e.\ with spacing $\Delta x= \Delta y= \Delta z$), with a marker-and-cell arrangement of velocity and pressure points, using a finite-difference method. All scalar fields are defined at the cell centers. Hereafter we describe the phase-change two-fluid solver, starting with the interface-capturing method.
\subsection{Interface representation}\label{num:int_repre}
To capture the interface, we use the MTHINC volume-of-fluid method \cite{ii2012interface}. For the sake of clarity, we start by briefly describing the original method and then explain our approach for modifying the advection scheme to account for a \emph{smooth but non-divergence-free} interface velocity, $\mathbf{u}_\Gamma$. The construction of the interface velocity is described later in section~\ref{num:int_vel}. \par 
We start by considering the cell-averaged volume fraction field, or volume-of-fluid function $C$, governed by the following equation:
\begin{equation}
    \dfrac{\partial C}{\partial t}+\nabla\cdot(\mathbf{u}_{\Gamma}H^{ht})=C\nabla\cdot\mathbf{u}_{\Gamma}\mathrm{,}
	\label{cadv}
\end{equation}
where $H^{ht}$ is a hyperbolic tangent function, approximating the phase indicator function $H$. The numerical fluxes are determined from the semi-analytical procedure described in \cite{ii2012interface}, without requiring an explicit interface reconstruction. \ns{The governing equation for $C$, eq.~\eqref{cadv}}, is integrated in time with a directional-splitting method \cite{puckett1997high,aulisa2003geometrical}, with an additional correction accounting for the non-zero divergence of the interface velocity. As described in \cite{ii2012interface}, the following implicit equations are solved sequentially from time step $n$ to $n+1$:
\begin{equation}
	\begin{aligned}[b]
		C_{i,j,k}^{*} &= C_{i,j,k}^{n}-\dfrac{1}{\Delta x}\left(f^{n}_{i+1/2,j,k}-f^{n}_{i-1/2,j,k}\right)+\dfrac{\Delta t}{\Delta x}C_{i,j,k}^{*}\left(u_{\Gamma i+1/2,j,k}^{n}-u_{\Gamma i-1/2,j,k}^{n}\right), \\
		C_{i,j,k}^{**} &= C_{i,j,k}^{*}-\dfrac{1}{\Delta y}\left(g^{*}_{i,j+1/2,k}-g^{*}_{i,j-1/2,k}\right)+\dfrac{\Delta t}{\Delta y}C_{i,j,k}^{**}\left(v_{\Gamma i,j+1/2,k}^{n}-v_{\Gamma i,j-1/2,k}^{n}\right), \\
		C_{i,j,k}^{***} &= C_{i,j,k}^{**}-\dfrac{1}{\Delta z}\left(h^{**}_{i,j,k+1/2}-h^{**}_{i,j,k-1/2}\right)+\dfrac{\Delta t}{\Delta z}C_{i,j,k}^{***}\left(w_{\Gamma i,j,k+1/2}^{n}-w_{\Gamma i,j,k-1/2}^{n}\right), \\
	\end{aligned}
	\label{fluxes}
\end{equation}
where $\Delta t$ is the time step, and $f^n$, $g^{*}$ and $h^{**}$ are the numerical fluxes computed as in~\cite{ii2012interface}, and $u_{\Gamma}$, $v_{\Gamma}$ and $w_{\Gamma}$ denote the three components of the interface velocity vector. Since each of the advection steps in eq.~\eqref{fluxes} corresponds to a non-divergence-free one-dimensional velocity field \cite{weymouth2010conservative}, the original method performs a further correction to ensure that the divergence-free condition is satisfied:
\begin{equation}
    C_{i,j,k}^{n+1}=C_{i,j,k}^{***}-\Delta tF_{i,j,k}^{n}\mathrm{,}
	\label{final_corr_1}
\end{equation}
where $F_{i,j,k}^n$ is given by:
\begin{align}
    F_{i,j,k}^n =\,\, &C_{i,j,k}^{*}\dfrac{u_{\Gamma i+1/2,j,k}^{n}-u_{\Gamma i-1/2,j,k}^{n}}{\Delta x}+\nonumber\\
                  &C_{i,j,k}^{**}\dfrac{v_{\Gamma i,j+1/2,k}^{n}-v_{\Gamma i,j-1/2,k}^{n}}{\Delta y}+\nonumber\\ 
                  &C_{i,j,k}^{***}\dfrac{w_{\Gamma i,j,k+1/2}^{n}-w_{\Gamma i,j,k-1/2}^{n}}{\Delta z}\mathrm{.}
\end{align}
The second term on the right-and-side of eq.~\eqref{final_corr_1} corresponds to the correction used in the conventional directional-splitting method for a divergence-free advection velocity. 
In the present work, we extend the directional-splitting advection method in eq.~\eqref{final_corr_1} to ensure that the corresponding non-zero interface velocity divergence is accurately prescribed:
\begin{equation}
    C_{i,j,k}^{n+1}=C_{i,j,k}^{***}-\Delta tF_{i,j,k}^{n}+\Delta t C_{i,j,k}^{n+1}\left(\nabla\cdot\mathbf{u}_\Gamma\right)_{i,j,k}^{n}\mathrm{.}
	\label{final_corr_2}
\end{equation}
The last term in eq.~\eqref{final_corr_2} can be seen as an implicit volume deflation step, and ensures that the correct value of the velocity divergence is used to update $C$ to time level $n+1$; $\left(\nabla\cdot\mathbf{u}_\Gamma\right)_{i,j,k}^{n}$ is the discrete divergence of $\mathbf{u}_{\Gamma}$ at time level $n$. \par
Once $C^{n+1}$ is determined, the thermophysical properties are updated using eq.~\eqref{material_prop}, where the volume fraction field $C$ is used as a smoothed approximation of the phase indicator function $H$. This makes all terms involving thermophysical properties numerically differentiable.

\subsection{Flow solver}\label{num:flow_solver}
The two-fluid Navier-Stokes solver uses a projection method \cite{chorin1968numerical} with the pressure-splitting technique described in \cite{dong2012time}, reducing the pressure-correction step to a constant-coefficients Poisson equation. The underlying idea is to split the variable-coefficients pressure Poisson equation into two parts: a variable-coefficients term that is treated explicitly by extrapolating the pressure field into the current time level, and a constant-coefficients term. This allows for using efficient FFT-based direct solvers for the second-order finite-difference Poisson equation with constant coefficients as shown in~\cite{dodd2014fast}, which are about one order of magnitude faster than a standard iterative solver. We should note however that the overall method can be easily applied to standard two-fluid solvers that do not use this pressure-splitting technique. \ns{The overall solution procedure uses an Adams-Bashforth method to advance the solution from time step $n$ to $n+1$, and is summarized below in semi-discrete form:
%\ns{a}
\begin{align}
    \hat{p}= \left(1+\frac{\Delta t^{n+1}}{\Delta t^n}\right)p^{n}-\left(\frac{\Delta t^{n+1}}{\Delta t^{n}}\right)p^{n-1}\mathrm{,}\\
    \rho^{n+1}\left(\dfrac{\mathbf{u}^\star-\mathbf{u}^{n}}{\Delta t^{n+1}}\right) = \left(1+\dfrac{1}{2}\dfrac{\Delta t^{n+1}}{\Delta t^n}\right)\mathbf{RHS}^{n}-\left(\dfrac{1}{2}\dfrac{\Delta t^{n+1}}{\Delta t^n}\right)\mathbf{RHS}^{n-1}\mathrm{,}\label{eqn:u_predic}\\
    \frac{\nabla^2p^{n+1}}{\rho_0} = \nabla\cdot\left[\left(\dfrac{1}{\rho_0}-\dfrac{1}{\rho^{n+1}}\right)\nabla\hat{p}\right]+ \nonumber\\ 
     \dfrac{1}{\Delta t^{n+1}}\left[\nabla\cdot\mathbf{u}^\star-\dot{m}^{n+1}\left(\dfrac{1}{\rho_2}-\dfrac{1}{\rho_1}\right)\delta_\Gamma^C\right],\label{eqn:lap_p}\\
	\mathbf{u}^{n+1} = \mathbf{u}^\star-\Delta t^{n+1}\left[\dfrac{1}{\rho_0}\nabla p^{n+1}+\left(\dfrac{1}{\rho^{n+1}}-\dfrac{1}{\rho_0}\right)\nabla\hat{p}\right],
\end{align}
where $\mathbf{u}^\star$ is the predicted velocity, $\Delta t^{n+1}$ and $\Delta t^{n}$ represent the time-step computed at time step $n+1$ and $n$ to fulfill the temporal stability requirements}, $\rho_0 = \min(\rho_1,\rho_2)$, $\mathbf{RHS}$ denotes the discretized advection, diffusion, gravity and surface tension terms in eq.~\eqref{mom}; $\delta_\Gamma^C=|\nabla C^{n+1}|$ is a regularized Dirac delta function approximating $\delta_{\Gamma}$ in eqs.~\eqref{mass},\eqref{mom} and \eqref{energy}, and is discretized using the Youngs method~\cite{youngs1984interface}. The continuum surface force model (CSF) proposed by Brackbill~\cite{brackbill1992continuum} is used for discretizing the surface tension term.\par
This set of equations is very close to that of \cite{dodd2014fast}, except for the last term of eq.~\eqref{eqn:lap_p}. One can easily see that this term approximates the right velocity divergence at the interface due to phase change, i.e.\ eq.~\eqref{mass}, regularized over the grid cells where $0<C<1$. \par
This CSF-like approach makes the final velocity field numerically differentiable. Still, the flow velocity $\mathbf{u}$ may vary strongly across the interface, and therefore the use of second-order central schemes for the spatial derivatives of the convective terms in $\mathbf{RHS}$ is not desirable. Accordingly, a QUICK scheme~\cite{leonard1979stable} is employed for the convective term \pc{(discretized as $\nabla\cdot(\mathbf{u}\otimes\mathbf{u})-\mathbf{u}\nabla\cdot\mathbf{u}$)}, while second-order central differences are used for the diffusion terms.\par
Our method uses the fast and versatile FFT-based DNS code \emph{CaNS} \cite{costa2018fft} as base Navier-Stokes solver. This solver allows for several combinations of homogeneous pressure boundary conditions, which is particularly convenient for the computational setups used in the present work. Finally, validations of the interface-capturing procedure in absence of phase change have been reported in~\cite{rosti2019numerical}, and its suitability for simulating complex and turbulent flows demonstrated in \cite{de2019effect,rosti_ge_jain_dodd_brandt_2019}.\par
\ns{It is worth remarking that the use of the pressure-splitting approach for solving the Poisson equation can be beneficial for interface resolved simulation of phase-changing problems, especially those involving high mass transfer rates between phases. In fact, as noted also in~\cite{sato2013sharp}, as $\dot{m}$ increases and as the resolution to properly resolve the governing equations is increased, the source term on the right-hand side of eq.~\eqref{eqn:lap_p} becomes more and more important, and might pose a stiffness problem. \pc{Hence, if a variable-coefficient pressure Poisson equation is solved using an iterative method, the convergence of the solver may be slow. This is not an issue in the present method, since our approach allows for a fast direct solver.}}

\subsection{Vapor mass transport}\label{num:vapor_sol}

Equation~\eqref{mass_fraction} is solved in $\Omega_2$ with the Dirichlet boundary condition prescribed at the interface in eq.~\eqref{eqn:bc_y}. This equation is discretized in time as follows:
\begin{equation}
	\dfrac{Y_2^{l,n+1}-Y_2^{l,n}}{\Delta t} = -\mathbf{u}^n\cdot\nabla Y_2^{l,n}+D_{lg}\nabla^2Y_2^{l,m},
	\label{mass_fraction_n}
\end{equation}
with $m=n+1$ or $n$, depending on whether the diffusion term is discretized implicitly or explicitly. The spatial discretization needs to be modified close to the interface to prescribe the boundary condition at $\mathbf{x} = \mathbf{x}_\Gamma$. To achieve this, the finite-difference stencil is modified in grid cells close to interface by constructing a signed distance (i.e.\ level-set) field $\phi$ from $C$, with the method proposed in \cite{russo2000remark,albadawi2013influence}. Here $\phi>0$ corresponds to $\Omega_1$, and $\phi<0$ to $\Omega_2$.\par
The advection term is discretized using an upwind scheme; see e.g.\ \cite{sato2013sharp}. Taking the discretization in $x$ as example, it reads:
\begin{equation}
	%\begin{aligned}
	\mathbf{u}\cdot\nabla Y_2^{l} = 
	\dfrac{(u_c+|u_c|)}{2}\left.\dfrac{\partial Y_2^{l}}{\partial x}\right|_{-}+\left.\dfrac{(u_c-|u_c|)}{2}\dfrac{\partial Y_2^{l}}{\partial x}\right|_{+}, %\\
	\label{conv}
\end{equation}
where $u_c$ is the $x$-velocity component interpolated into the cell center (i.e.\ $u_c=(u^{c}_{i+1/2,j,k}+u^{c}_{i-1/2,j,k})/2$). When the interface crosses a grid cell, the discretized form of the gradients of $Y_2^l$ should be modified to conform to the interface boundary condition. This is achieved by considering a higher-order one-sided difference on an irregular stencil. For instance, the gradient of $Y_2^l$ at $i-1/2$ is computed as follows (the procedure for $i+1/2$ is analogous):
\begin{equation}
	\left.\dfrac{\partial Y_2^{l}}{\partial x}\right|_{-} = \begin{cases}
        \beta_0Y_{2,\Gamma}^{l}(T_{\Gamma,x}^{-})+\displaystyle{\sum_{p=0}^{3}\beta_{p+1}Y_{2,i+p}^{l}}  &\text{if $\phi_{i-1}\phi_{i}<0$}, \\
	\dfrac{Y_{2,i}^{l}-Y_{2,i-1}^{l}}{\Delta x} &\text{otherwise}\mathrm{,}
	\label{yvm}
\end{cases}
\end{equation}
where $Y_{2,\Gamma}^{l}$ is computed from eqs.~\eqref{eqn:bc_y} and~\eqref{eqn:eq_state}, with the interface temperature $T_{\Gamma,x}^{-}$ estimated from the neighboring values of $\phi$ \cite{liu2000boundary,tanguy2007level}:
\begin{equation}
	T_{\Gamma,x}^{-}=\dfrac{T_{i-1,j,k}|\phi_{i,j,k}|+T_{i,j,k}|\phi_{i-1,j,k}|}{|\phi_{i-1,j,k}|+|\phi_{i,j,k}|}.
	\label{tmp_int}
\end{equation}
In eq.~\eqref{yvm}, the one-sided difference coefficients $\beta$ are computed following the approach reported in~\cite{fornberg1988generation}, as in~\cite{dodd2017direct}. The resulting stencil is given by $x=x_i+\{-\theta_{x}^{-},0,1,2,3\}\Delta x$. Using the level-set function, the coefficient $\theta_{x}^{-}=(x_i-x_\Gamma)/\Delta x$ is computed as proposed in~\cite{liu2000boundary}:
\begin{equation}
    \theta_{x}^{-} = \dfrac{|\phi_{i,j,k}|}{|\phi_{i-1,j,k}|+|\phi_{i,j,k}|}.
\end{equation}
In case of small values of $\theta_{x}^{-}<0.25$, the point $x_i$ is removed from the one-sided difference stencil to prevent errors as it approaches a singular value~\cite{dodd2017direct}. Finally, the above procedure is performed in a dimension-by-dimension manner for directions $y$ and $z$.\par
\ns{Note that the aforementioned approach to discretize the vapor mass equation is not sufficiently robust when complex topological changes such coalescence and merging occur. Equation~\eqref{yvm} requires a stencil of three grid cells in the vapor domain and, therefore, when two or multiple droplets start to coalesce and merge, numerical problems occur since there are not enough vapor grid cells for an accurate estimation of the gradient of $Y_{2}^l$. In these cases, we decide to reduce the stencil of equation~\eqref{yvm} such that $x=x_i+\{-\theta_{x}^{-},0,1\}\Delta x$ and compute $\beta$ accordingly. Even though this approach reduces the accuracy of the calculation of the vapor mass gradient, we show in the Results Section (case~\ref{subsec:case5}) that it still provides stable and grid convergent results.} \par
If an explicit temporal discretization of eq.~\eqref{mass_fraction_n} is employed, the second derivatives in the diffusion term are discretized using the first derivatives, previously computed. Taking once more the term in $x$ as an example, the second derivative reads~\cite{gibou2002second}:
\begin{equation}
	\dfrac{\partial^2 Y_2^{l}}{\partial x^2} = \dfrac{1}{\Delta x}\left(\left.\dfrac{\partial Y_2^{l}}{\partial x}\right|_{+}-\left.\dfrac{\partial Y_2^{l}}{\partial x}\right|_{-}\right)\mathrm{.}
	\label{second_dev}
\end{equation}
    Conversely, if an implicit discretization is chosen, the procedure involves solving an Helmholtz equation where the boundary condition in eq.~\eqref{eqn:bc_y} is prescribed with the aid of the constructed level-set field~\cite{gibou2007level,tanguy2007level}. The resulting symmetric definite positive linear system is solved with the parallel semicoarsening multigrid solver (PFMG) of the HYPRE library~\cite{falgout2006design}.
\subsection{Interfacial vapor mass flux}\label{num:mass_calc}
The calculation of the interfacial vapor mass flux is accomplished using eq.~\eqref{mass_flux}, written in terms of $\dot{m}$:
\begin{equation}
    \dot{m}=-\dfrac{\rho_2D_{lg}}{1-Y_{2,\Gamma}^{l}}\nabla_\Gamma Y_2^l\cdot\mathbf{n}.
	\label{eqn:mass_flux_2}
\end{equation}
Since $Y_2^l$ is not defined in $\Omega_1$, standard finite differences cannot be directly used to approximate the gradient at the interface. Instead, we follow the approach firstly proposed in~\cite{aslam2004partial} and used in e.g.\ \cite{tanguy2014benchmarks} and extend the $Y_2^l$ into $\Omega_1$. Hence, before computing $\dot{m}$ from eq.~\eqref{eqn:mass_flux_2}, $Y_2^l$ is extrapolated into the liquid domain using a second-order pde-based extrapolation. This involves the successive solution to steady state of the three following equations:
\begin{equation}
	\begin{aligned}
        \dfrac{\partial Y_{2,\mathrm{nn}}^{l,e}}{\partial\tau}+H_{\phi}(\phi+a_2\Delta l)\mathbf{n}\cdot\nabla Y_{2,\mathrm{nn}}^{l,e}=0\mathrm{,} \\
        \dfrac{\partial Y_{2,\mathrm{n}}^{l,e}}{\partial\tau}+H_{\phi}(\phi+a_1\Delta l)(\mathbf{n}\cdot\nabla Y_{2,\mathrm{n}}^{l,e}-Y_{2,\mathrm{nn}}^{l,e})=0\mathrm{,} \\
        \dfrac{\partial Y_{2}^{l,e}}{\partial\tau}+H_{\phi}(\phi+a_0\Delta l)(\mathbf{n}\cdot\nabla Y_{2}^{l,e}-Y_{2,\mathrm{n}}^{l,e})=0\mathrm{,}
	\end{aligned}
	\label{eqn:extrapl}
\end{equation}
where $Y_{2,\mathrm{n}}^{l,e}$ and $Y_{2,\mathrm{nn}}^{l,e}$ are the first and the second derivative of $Y_{2}^{l}$ along the $\Omega_1$-pointing interface normal, $H_{\phi}$ represents a cosine regularized Heaviside function~\cite{sussman1998improved,son2008numerical}, computed from the level-set function $\phi$ and $\tau$ is a pseudo time used only to advance to the steady state solution eqs.~\eqref{eqn:extrapl}. %
The offsets $a_i\Delta l$ (with $\Delta l$ being the grid spacing) ensure that only values in the vapor domain, i.e.\ at $\phi<0$ are used to solve eqs.~\eqref{eqn:extrapl}. When implicit diffusion is used to discretize eq.~\eqref{mass_fraction_n}, the values suggested in the original reference suffice for an accurate extrapolation of the vapor field onto the liquid domain. However, if this equation is discretized explicitly, care should be taken when the phase in a certain grid cell changes (e.g.\ from liquid to gas). If the solution is not allowed to gradually adapt itself (within a few time steps) to the sudden change in phase, numerical errors may occur. For this reason, in this case, the solution  is extrapolated from a position slightly inward of the interface. The corresponding shifted offset is now given by $(a_i+1)\Delta l$ (i.e.\ one grid size inward of the interface).\par
\pc{It is worth noting that modifications in the computation of derivatives of mass fraction in the extrapolation procedure are needed e.g.\ when two or more droplets are too close to each other. When the finite-difference stencil for the computation of a derivative at two liquid grid cells separated a thin gas layer, the spatial derivatives of the mass fraction are approximated as follows:
\begin{equation}
	\dfrac{\partial Y_{2,\mathrm{n}}^l}{\partial x} = \begin{cases}
        \dfrac{Y_{2,\Gamma}^{l}(T_{\Gamma}^{+})-Y_{2,\Gamma}^{l}(T_{\Gamma}^{-})}{|x_\Gamma^+-x_\Gamma^-|}  &\text{if $\varepsilon\Delta x<|x_\Gamma^+-x_\Gamma^-|<\Delta x$}, \\
	0 &\text{if $|x_\Gamma^+-x_\Gamma^-|<\varepsilon\Delta x$}\mathrm{,}
	\label{yvm_nor}
\end{cases}
\end{equation}
where we set $\varepsilon=0.25$; $T_{\Gamma}^{+}$ and $T_{\Gamma}^{-}$ are computed using equation~\eqref{tmp_int}, and $x_\Gamma^+$ and $x_\Gamma^-$ represent the two interface locations computed using $\phi_{i+1,j}$ and $\phi_{i-1,j}$. This is to say that if two liquid interfaces separated by a thin layer of gas with thickness smaller than $0.25\Delta x$, we consider the gradient to be under-resolved and set it to zero.
} \par

\ns{Eqs.~\eqref{eqn:extrapl} are integrated in time using about $15$ forward Euler pseudo time steps, and a first-order upwind scheme \cite{tanguy2014benchmarks} to advect the scalar fields. The resulting extended vapor species field $Y_2^{l,e}$ is then used to compute the gradients in eq.~\eqref{eqn:mass_flux_2} using a central difference scheme. Moreover, in the calculation of the interfacial mass flux we use the \nf{unit normal vector computed from the VoF field directly from its definition, i.e., $\mathbf{n}=\nabla C/|\nabla C|$. As done for the regularized Dirac delta function in eqs.~\eqref{mass},\eqref{mom} and \eqref{energy}, the gradients of the color function are evaluated using finite differences, following the Youngs approach~\cite{youngs1982time}. Accordingly, $\mathbf{n}$ can be directly employed in eq.~\eqref{eqn:mass_flux_2} and in the calculation of the interface velocity (see section~\ref{num:int_vel}).}} \par
\pc{The interfacial values required for computing the mass flux in eq.~\eqref{eqn:mass_flux_2} are not extended in a band around the interface. Instead, we take the simple approach of approximating these quantities by the local extrapolated field quantities, i.e.\ $Y_{2,\Gamma}^l\approx Y_{2}^{l,e}$, and $\nabla_\Gamma Y_2^l\cdot\mathbf{n}\approx \nabla Y_2^{l,e}\cdot\mathbf{n}$ \cite{tanguy2007level,villegas2016ghost}. Potential improvements of this approach are a constant extrapolation of the interfacial mass flux to a band around the interface \cite{tanguy2014benchmarks}, or the approach in \cite{chai2018coupled} that extrapolates the interface temperature to a band around the interface, and computes $Y_{2,\Gamma}^l$ from this extrapolated temperature field. We have tested these approaches and observed a marginal improvement in the accuracy of mass conservation for the cases presented in this manuscript.}\par
\ns{Overall, the approach described in this section allows us to define $\dot{m}$ in a band of about $6\, \Delta l$ around the interface, which covers the region where $0<C<1$.}
\subsection{Energy equation}\label{num:en_eqn}
\ns{All the terms in the energy equation~\eqref{energy} are discretized explicitly using an Adams-Bashforth scheme whose coefficients are computed for a variable time step:
\begin{align}
    \dfrac{T^{n+1}-T^n}{\Delta t^{n+1}} = \left(1+\dfrac{1}{2}\frac{\Delta^{n+1}}{\Delta t^{n}}\right)\left(-\mathbf{u}\cdot\nabla T+\dfrac{1}{\rho c_p}\mathrm{RHS}_T\right)^{n} \nonumber \\ -\left(\dfrac{1}{2}\dfrac{\Delta t^{n+1}}{\Delta t^n}\right)\left(-\mathbf{u}\cdot\nabla T+\dfrac{1}{\rho c_p}\mathrm{RHS}_T\right)^{n-1}
\mathrm{,}
	\label{eqn:energy_disc}
\end{align}
}
%\par
where $\mathrm{RHS}_T$ denotes the discretized form of the diffusive and source terms in eq.~\eqref{energy}\pc{, and the temperature field around the interface is used as an approximation of $T_{\Gamma}$ in the source term.} \ns{Note again that, as specified after eq.~\eqref{material_prop}, we employ \nf{the harmonic average for the term $\rho c_p$ to perform a consistent discretization of the diffusion and jump terms, as eq.~\eqref{eqn:energy_disc} suggests. In fact, we find that the use of the arithmetic mean may lead to unphysical values in the temperature field.}} Our spatial discretization uses the $5$th-order WENO scheme described in~\cite{castro2011high}, while the diffusion term is treated using standard central differences. The Dirac delta function in the singular term is regularized as explained above for the discretization of the momentum equations.\par
\ns{Finally, it is worth noting that eq.~\eqref{eqn:energy_disc} is discretized in non-conservative form, which becomes more and more inaccurate as both the density and specific heat capacity ratio increase. \pc{We have therefore compared the numerical results employing the conservative and the non-conservative discretization of the energy equation for a vaporizing droplet in the Results Section (case~\ref{subsec:case3}). Indeed, we for low resolutions the conservative scheme shows better agreement with the reference data. Nevertheless, as the grid is refined the results converge to the same numerical value, regardless of the formulation.}}

\subsection{Interface velocity construction}\label{num:int_vel}
The calculation of the interface velocity $\mathbf{u}_{\Gamma}$ for the advection of $C$ is a key aspect of this method. The main challenge stems from the discontinuity of the one-fluid-formulation velocity $\mathbf{u}$ across the interface, while $\mathbf{u}_\Gamma$ as defined in eq.~\eqref{rankine} is continuous. \par
To overcome this issue, we adopt a strategy that accurately extends the liquid velocity into the gas domain, making the extended field differentiable in $\Omega$. The basic idea is that phase change induces a Stefan flow, which is responsible for the jump in the flow velocity $\mathbf{u}$. Hence, a divergence-free liquid-velocity extension can be obtained by subtracting this jump to $\mathbf{u}$. The Stefan flow velocity $\mathbf{u}^S$ can be obtained from a velocity potential $\varphi$ as follows:
\begin{equation}
	\begin{cases}
		\nabla^2\varphi &= \dot{m}\left(\dfrac{1}{\rho_2}-\dfrac{1}{\rho_1}\right)|\nabla C|\mathrm{,} \\
		\mathbf{u}^S &= \nabla\varphi\mathrm{.}	
	\end{cases}
	\label{ext_div}
\end{equation}
The solution of the first equation in~\eqref{ext_div} can be, once more, obtained for a wide variety of boundary conditions for $\varphi$ using the efficient FFT-based direct solver described in \cite{costa2018fft}. The main advantage of this approach is that, since $\mathbf{u}^S$ is defined in the entire domain, an extended liquid velocity $\mathbf{u}_{1}^{e}$ can be easily computed:
\begin{equation}
	\mathbf{u}_{1}^e = \mathbf{u}-\mathbf{u}^{S}.
	\label{jump_subtr}
\end{equation}
\ns{It can be easily shown that $\mathbf{u}_{1}^{e}$ is divergence-free by construction. Moreover, for the same reason mentioned in section~\ref{num:flow_solver}, the possibility to use direct solvers for eq.~\eqref{ext_div} allows to circumvent potential stiffness problems induced by high values of the interfacial mass flux.} \par
Note that a similar idea has been recently developed in~\cite{malan2018direct} for boiling flows, by solving for \ns{an extended velocity derived from a density-weighted variable coefficient Poisson equation, reported here for completeness:
\begin{equation}
	\begin{cases}
		\nabla\cdot\left(\dfrac{1}{\rho}\nabla\varphi\right) &= \dot{m}\left(\dfrac{1}{\rho_2}-\dfrac{1}{\rho_1}\right)|\nabla C|\mathrm{,} \\
		\mathbf{u}^S &= \dfrac{1}{\rho}\nabla\varphi\mathrm{.}	
	\end{cases}
	\label{ext_div_var_coeff}
\end{equation}
Although equation~\eqref{ext_div_var_coeff} directly accounts for the density contrast between the two phases \pc{like the variable-density pressure Poisson equation}, it has the disadvantage of requiring the solution of a variable-coefficient Poisson equation that cannot be solved using fast FFT-based direct solvers.} \par
\ns{In a similar way to what it is commonly done in the prediction/correction procedure for the Navier-Stokes solver, the boundary conditions for $\varphi$ and $\mathbf{u}_1^e$ have to be prescribed consistently, \pc{and are set to be the same as those of the pressure and prediction velocity, respectively.}}\par
Finally, eq.~\eqref{rankine} can be used to compute $\mathbf{u}_{\Gamma}$:
\begin{equation}
    \mathbf{u}_{\Gamma} = \mathbf{u}_{1}^{e}-\dfrac{\dot{m}}{\rho_1}\mathbf{n}.
	\label{eqn:int_vel}
\end{equation}
\par
The interface velocity $\mathbf{u}_{\Gamma}$ is then used to advect $C$, as described in section~\ref{num:int_repre}. Furthermore, it important to stress that following this approach both $\mathbf{u}_1^e$ and $\mathbf{u}_{\Gamma}$ are equal to the one-fluid velocity derived from the Navier-Stokes solver $\mathbf{u}$ in those region away from the interface where $\dot{m}$ is not defined. \ns{Note that in principle the procedure here described can be applied also to those interface capturing methods based e.g.\ on level-set and conservative level-set methods while maintaining the whole domain formulation for solving the momentum equations.} Few words should be also spent about the overhead associated with the computation of $\mathbf{u}_{\Gamma}$. For a three-dimensional case (evaluated for the last setup described in section~\ref{subsec:case6}), we found that it is relatively small, i.e. less than $2\%$ of the total wall-clock time per time step.
\par 
We should note that $C$ may also be advected directly using $\mathbf{u}_{1}^{e}$ and a source term accounting for phase change on the right-hand-side. Inserting eq.~\eqref{eqn:int_vel} into \eqref{cadv}, we get:
\begin{equation}
    \dfrac{\partial C}{\partial t}+\nabla\cdot(\mathbf{u}_1^eH^{ht})=-\dfrac{\dot{m}}{\rho_1}|\nabla C|\mathrm{.}
\end{equation}
Since $\mathbf{u}_{1}^{e}$ is divergence-free, the original splitting advection in \cite{weymouth2010conservative} can be used, followed by a step that accounts for the source term \cite{malan2018direct}. Test simulations reproducing the benchmarks in section~\ref{sec:results} showed that \pc{the circular shape of a vaporizing droplet in a quiescent medium} is better preserved \pc{(lower errors in the perimeter of the droplet)} when the phase-change term is explicitly accounted for in the split advection steps (eq.~\ref{fluxes}), i.e.\ when $\mathbf{u}_\Gamma$ is explicitly used to advect the interface as described above in section~\ref{num:int_repre}.

\subsection{\ns{Overall solution procedure} and time marching}
\ns{We briefly summarize the overall solution procedure, which is also outlined in figure~\ref{fig:flowchart} for clarity. First, the volume-of-fluid function is advanced to $C^{n+1}$, the corresponding interface normal vector and curvature are updated, and the level-set is reconstructed. Next, the vapor mass-fraction equation is solved to obtain $Y_2^{l,n+1}$, and the interfacial mass flux $\dot{m}^{n+1}$ is computed. Then the energy and Navier-Stokes equations can be advanced in time to obtain $\mathbf{u}^{n+1}$, $p^{n+1}$ and $T^{n+1}$. Finally, the time marching ends with the computation of the interface velocity $\mathbf{u}_\Gamma^{n+1}$, which is used to transport $C$ at the next time level.\par
\begin{figure}[t!]
    \centering
    \includegraphics[width=0.85\textwidth]{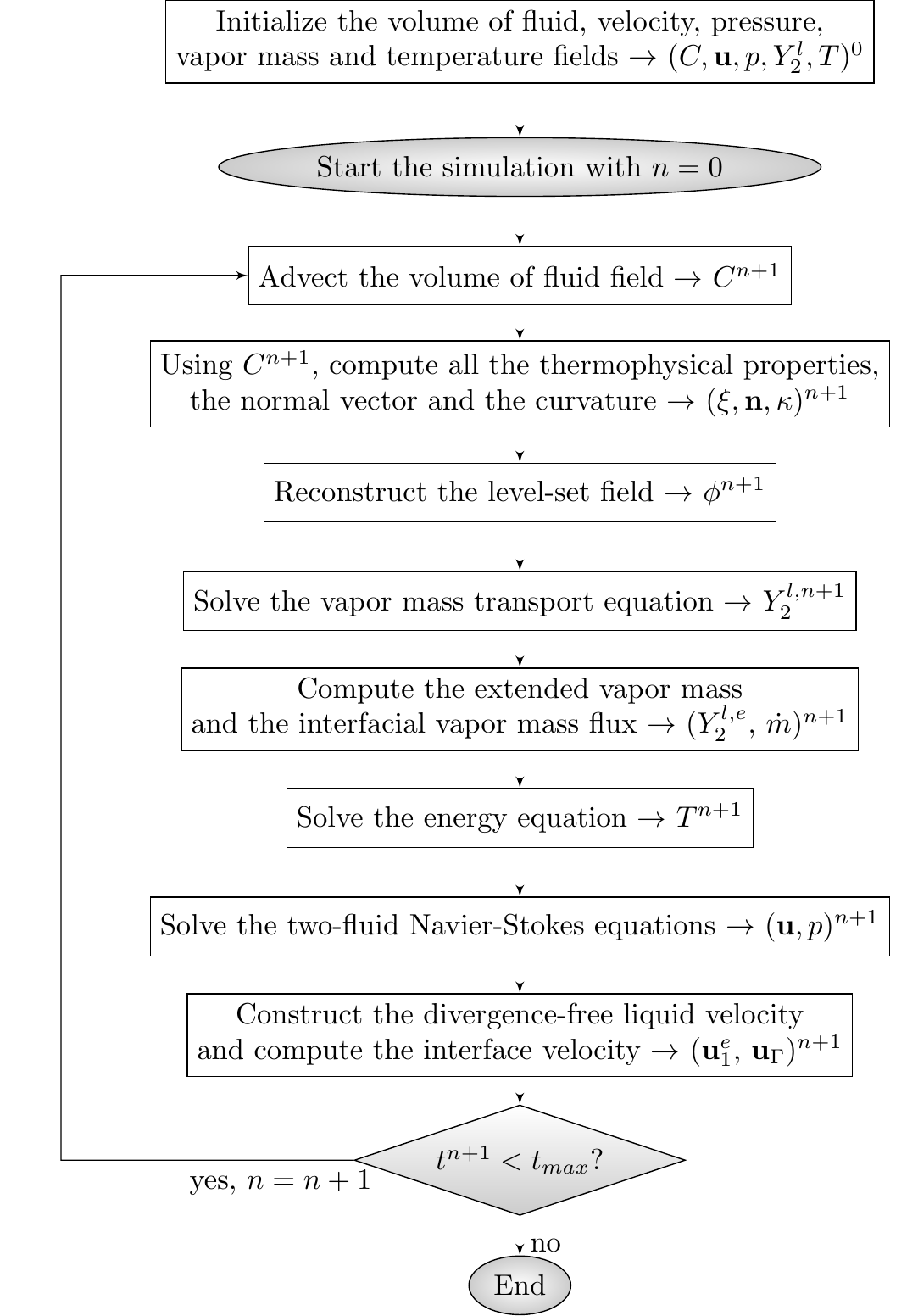}
    \caption{\pc{Diagram illustrating the overall solution procedure. $t^{n+1}$ is the physical corresponding to time-step $n+1$ and $t_{max}$ is the maximum physical time reached in the simulation.}}
	\label{fig:flowchart}
\end{figure}
The time step $\Delta t^{n+1}$ is estimated from the stability constraints of the overall system:}
\begin{equation}
	\Delta t^{n+1}=C_{\Delta t}\min(\Delta t_c,\Delta t_{\sigma},\Delta t_{\mu},\Delta t_m,\Delta t_e)^{n+1}\mathrm{,}
	\label{eqn:max_dt}
\end{equation}
where $\Delta t_c$, $\Delta t_{\sigma}$, $\Delta t_{\mu}$, $\Delta t_{m}$ and $\Delta t_e$ are the maximum allowable time steps due to convection, surface tension, momentum diffusion, vapor mass diffusion and thermal energy diffusion. These are determined as suggested in~\cite{kang2000boundary}:
\begin{equation}
	\begin{aligned}
	\Delta t_c & =\left(\dfrac{|u_{x,\max}|}{\Delta x}+\dfrac{|u_{y,\max}|}{\Delta y}+\dfrac{|u_{z,\max}|}{\Delta z}\right)^{-1}\mathrm{,} \\
	\Delta t_{\sigma}&=\sqrt{\dfrac{(\rho_1+\rho_2)\min(\Delta x^3,\Delta y^3,\Delta z^3)}{4\pi\sigma}}\mathrm{,} \\
	\Delta t_{\mu}&=\left[\max\left(\dfrac{\mu_1}{\rho_1},\dfrac{\mu_2}{\rho_2}\right)\left(\dfrac{2}{\Delta x^2}+\dfrac{2}{\Delta y^2}+\dfrac{2}{\Delta z^2}\right)\right]^{-1}\mathrm{,} \\
	\Delta t_{m}&=\left[D_{lg}\left(\dfrac{2}{\Delta x^2}+\dfrac{2}{\Delta y^2}+\dfrac{2}{\Delta z^2}\right)\right]^{-1}\mathrm{,} \\ 
	\Delta t_{e}&=\left[\max\left(\dfrac{k_1}{\rho_1c_{p,1}},\dfrac{k_2}{\rho_2c_{p,2}}\right)\left(\dfrac{2}{\Delta x^2}+\dfrac{2}{\Delta y^2}+\dfrac{2}{\Delta z^2}\right)\right]^{-1}\mathrm{,}
	\end{aligned}
	\label{eqn:diff_con_t}
\end{equation}
where $|u_{i,\max}|$ is an estimate of the maximum value of the $i$th component of the flow velocity; $\Delta t_m$ is only considered when the vapor mass diffusion term is discretized explicitly. Setting $C_{\Delta t}=0.35$ in the present work was seen to be sufficient for a stable and accurate time integration.\par
Finally, we should note that our framework can be easily adapted to a two-phase flow undergoing temperature-induced phase change in a single-component system, i.e.\ boiling. This procedure is described in \ref{sec:appendixA}.

\section{Results}\label{sec:results}
%{I saw that many authors use "Results and discussion", is it okay?} \par 
%since it is a numerical paper, and we mostly validate the method, we can just call it results
We present now a validation of our method against several benchmark cases. For clarity, the physical parameters defining the different setups are displayed in table~\ref{tbl:table_groups}. Unless otherwise stated, the time step is set to be constant and determined from eq.~\eqref{eqn:max_dt} with $C_{\Delta t} = 0.35$, and the diffusion term in eq.~\eqref{mass_fraction_n} governing $Y_2^l$ is discretized implicitly.
\begin{sidewaystable}%[h!]
    \centering
	\begin{center}
	\end{center}
		\begin{tabular}{c|cccccccccccccc}
	    	\toprule
            Section & $\mathrm{Re}$ & $\mathrm{We}$ &  $\mathrm{Fr}$ & $\mathrm{Pr}$ & $\mathrm{Sc}$ & $\mathrm{Ste}$ & $\mathrm{Ste}^m$ & $\lambda_{\rho}$ & $\lambda_{\mu}$ &  $\lambda_{c_p}$ & $\lambda_k$ & $	\lambda_M$ & $u_{ref}$ & $T_{ref}$ \\
 		   	\midrule
                \ref{subsec:case1}  & $~25$ & $0.1$ & $\infty$ & N/A   & N/A     & N/A    & N/A    & $10-100$ & $50$  & N/A    & N/A    & N/A    & $\dot{m}_0/\rho_2$ & N/A \\
                \ref{subsec:case2}  & $~25$ & $0.1$ & $\infty$ & N/A   & $0.04$ & N/A    & N/A    & $10-100$ & $50$  & N/A    & N/A    & N/A    & $D_{lg}/d_0$ & N/A  \\
                \ref{subsec:case3}  & $~1.50$ & $3.31\cdot 10^{-5}$ & $\infty$ & $0.69$ & $0.67    $ & $\text{varied}$ & $0.07$ & $8.33$     & $63.68$  & $4.16$ & $0.23$ & $0.62$ & $D_{lg}/d_0$ & $T^{sat}-T_\infty$  \\
                \ref{subsec:case4}  & $~8.84$ & $3.125~$ & $1$      & $0.7$ & $1    $ & $\text{varied}$ & $0.07$ & $10$     & $50$  & $4.0~$ & $2.0~$ & $0.62$ & $\sqrt{|\mathbf{g}|d_0}$ & $T^{sat}-T_\infty$ \\
                \ref{subsec:case5}  & $25$ & $0.7$ & $\infty$      & $0.7$ & $0.035$ & $0.17$ & $0.0175$ & $10$     & $50$  & $4.0~$ & $2.0~$ & $0.62$ & $u_{r,0}$ & $T^{sat}-T_\infty$ \\
                \ref{subsec:case6}  & $~90$ & $0.5$ & $1$      & $0.7$ & $0.065$ & $0.08$ & $0.75$ & $10$     & $50$  & $4.0~$ & $2.5~$ & $0.62$ & $\sqrt{|\mathbf{g}|d_0}$ & $T^{sat}-T_\infty$\\
    		\bottomrule
    	\end{tabular}
    \caption{Governing parameters for the cases presented in this section. Their definitions are given in section~\ref{sec:physical_param}. For all cases $l_{ref}=d_0$, where $d_0$ is the initial droplet diameter.}
	\label{tbl:table_groups}
    %\centering
\end{sidewaystable}
\subsection{Droplet evaporation due to a prescribed, constant mass flux}\label{subsec:case1}
This case considers a phase-changing two-dimensional circular droplet, where evaporation is driven by a constant mass flux $\dot{m}_0$. This simple configuration allows us to verify the numerical method for phase-changing two-fluid flows, decoupled from the transport equations of energy and vapor mass. Under these conditions, it is easy to show that the droplet diameter evolves in time as follows~\cite{tanguy2007level}:
\begin{equation}
    \dfrac{d(t)}{d_0}=1-\left(\dfrac{2\dot{m}_0}{d_0\rho_1}\right)t\mathrm{.}\label{eqn:diam_cmflux}
\end{equation}\par
The circular droplet has an initial diameter $d_0$, it is centered in a square domain with dimensions $[-2\, d_0,2\, d_0]^2$, and zero-pressure outflow boundaries. The corresponding physical parameters have been reported in table~\ref{tbl:table_groups}. We are in particular interested in assessing the ability of the method to handle interface-normal velocity jumps across $\Omega$ (see eq.~\ref{rankine}). In order to test different magnitudes of this jump, we consider three density ratios, $\lambda_\rho = \lbrace 10,\, 50,\, 100 \rbrace$ while keeping the other parameters fixed.\par
\begin{figure*}[t!]
    \centering
    \begin{subfigure}[t]{0.5\textwidth}
        \centering
        \includegraphics[width=\textwidth,height=5.2 cm]{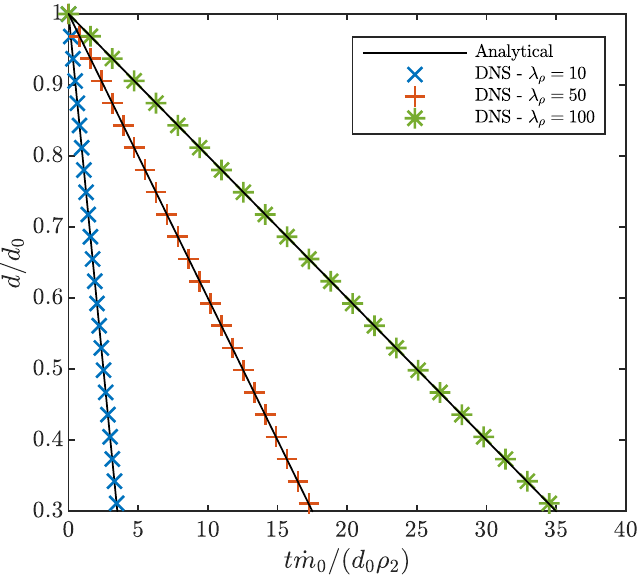}
        %\caption{}
    \end{subfigure}%
    ~ 
    \begin{subfigure}[t]{0.5\textwidth}
        \centering
        \includegraphics[width=\textwidth,height=5.2 cm]{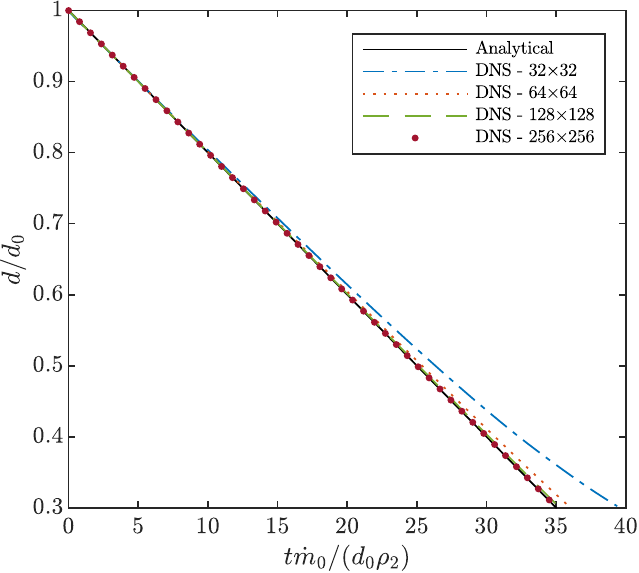}
        %\caption{}
    \end{subfigure}
    \put(-190,0){\small(\textit{a})}
    \put( -10,0){\small(\textit{b})}
    \caption{(\textit{a}): temporal evolution of the droplet diameter for $\lambda_{\rho}=10,50,100$, on a $256\times 256$ grid. (\textit{b}): grid convergence test for an evaporating droplet with $\lambda=100$.}
	\label{fig:eva_con_rdiff}
\end{figure*}
\begin{figure}[t!]
	\centering
	\includegraphics[width=\textwidth]{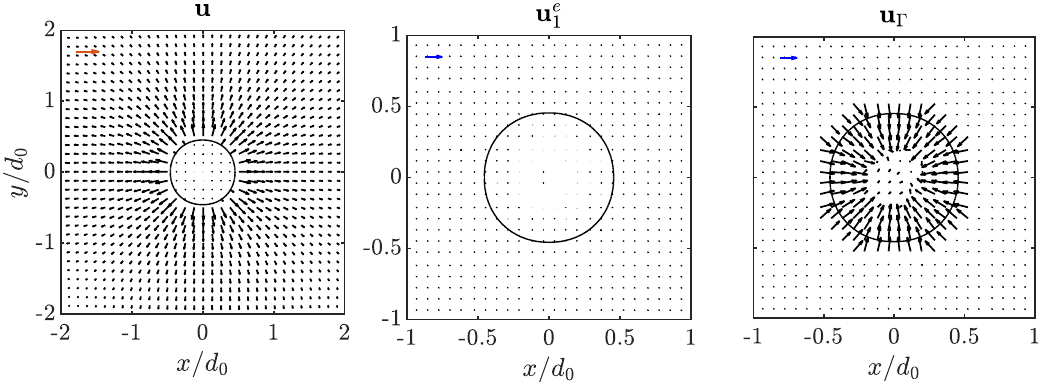}
    \put(-240,0){\small(\textit{a})}
    \put(-125,0){\small(\textit{b})}
    \put( -10,0){\small(\textit{c})}
    \caption{Vector field of (\textit{a}) flow velocity $\mathbf{u}$, (\textit{b}) divergence-free extended liquid velocity $\mathbf{u}_1^e$, and (\textit{c}) interface velocity $\mathbf{u}_{\Gamma}$, at $d/d_0=0.9$ for uniform evaporation on a $256\times 256$ grid and $\lambda_{\rho}=100$. The interface is represented by the solid black line. The vector scale (see top-left reference arrows) is equal to $||\mathbf{u}||_{\max}=\dot{m}(1/\rho_2-1/\rho_1)$ for panel (\textit{a}), and to $0.01||\mathbf{u}||_{\max}$ for panels (\textit{b}) and (\textit{c}). \ns{Panels \textit{b} and \textit{c} zoom into the domain $[-d_0,\, d_0]^2$} and the values of $\mathbf{u}_1^e$ and $\mathbf{u}_{\Gamma}$ outside the range where $\min(C,1-C)<10^{-8}$ have been clipped.}%\pc{wait for the final figure to write more detail on the scaling of b and c.}}
	\label{fig:eva3_case1}
\end{figure}

Figure~\ref{fig:eva_con_rdiff}(\textit{a}) shows the time evolution of the normalized droplet diameter, computed from its volume, for different $\lambda_\rho$, on the finest grid considered ($256\times 256$). The numerical results show excellent agreement with the analytical solution in eq.~\eqref{eqn:diam_cmflux}.
Panel~(\textit{b}) of the same figure shows the solution grid convergence for the case with the highest velocity jump, $\lambda_\rho=100$. The results illustrate a very good agreement even for relatively coarse grids. This result is expected from how the interface velocity $\mathbf{u}_{\Gamma}$ is constructed (eq.~\eqref{eqn:int_vel}). We recall that the volume deflation term in eq.~\eqref{final_corr_2}, controlling the bulk value of $C$, is proportional to the interface velocity divergence. 
The first term contributing to $\mathbf{u}_{\Gamma}$ in eq.~\eqref{eqn:int_vel} is a divergence-free extension of the liquid velocity, which effectively conserves the total volume-of-fluid to machine precision. The second term contains the interfacial mass flux and the interface normal. Since $\dot{m}$ is constant in this example, the only source of numerical error in the droplet volume is the divergence of the interface normal vector, i.e.\ the local curvature.
For an evaporating static droplet, the liquid velocity should be zero. We therefore expect the divergence-free liquid velocity extension to be $\mathbf{u}_1^e = 0$, and thus $\mathbf{u}_{\Gamma} =- (\dot{m}_0/\rho_1)\mathbf{n}$. This is illustrated in figure~\ref{fig:eva3_case1}, where we display the vector field $\mathbf{u}$ in panel (\textit{a}), $\mathbf{u}_1^e$ in (\textit{b}) and $\mathbf{u}_{\Gamma}$ in  (\textit{c}) for a physical time corresponding to $d/d_0=0.9$ and $\lambda_{\rho}=100$. Panel (\textit{a}) shows the expected velocity field $\mathbf{u}$, with a clear jump across $\Gamma$. The liquid velocity extension shown in panel (\textit{b}) is, expectedly, very small, with spurious velocities about three orders of magnitude smaller than the maximum value of $\mathbf{u}$. Finally, the interface velocity field shows the expected values for an evaporating droplet (note that $\mathbf{n}$ is only defined where $0<C<1$, and so is $\mathbf{u}_{\Gamma}$).
\par
\begin{figure*}[t!]
        \centering
        \includegraphics[width=6.4 cm,height=5.0 cm]{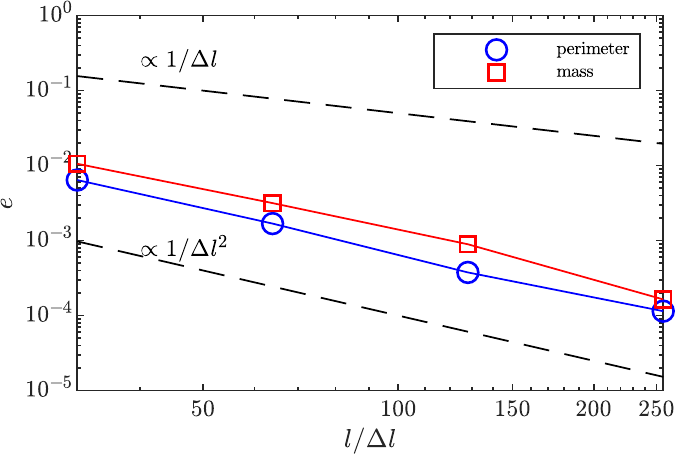}
    \caption{Error in the normalized droplet mass, $M/M_{t=0}$, and perimeter, $P/P_{t=0}$ for $\lambda_{\rho}=100$ and $d/d_0=0.4$. $\Delta l$ and $l$ denote the grid spacing and domain length.}
	\label{fig:order_100_case1}
\end{figure*}

Finally, we analyze the accuracy of our method by inspecting the convergence of the droplet mass and shape errors with increasing resolution. We consider the case with most significant velocity jump, $\lambda_\rho=100$ and a time step $\Delta t=0.0075\, d_0\rho_2/\dot{m}_0$, sufficiently low for errors in the temporal discretization to be negligible. Moreover, we ensure that the initial condition for $C$ yielded the same total droplet mass for all grids. 
The shape accuracy is assessed by computing the perimeter from the discrete integral of $|\nabla C|$ over the entire domain \nf{as follows (for a two-dimensional configuration):
\begin{equation}
	P=\sum_{j=1}^{N_y}\sum_{i=1}^{N_x}|\nabla C|_{i,j}\Delta x \Delta y\mathrm{,}
\end{equation}
where $N_x$ and $N_y$ are the number of grid points along the $x$ and $y$ direction.} Figure~\ref{fig:order_100_case1} reports the relative error in mass and perimeter, at physical time $t$ for which $d(t)/d_0=0.4$, \ns{i.e.:
\begin{equation}
  e(\Delta l,t) = \dfrac{\left|\chi_n(\Delta l,t)-\chi_a(\Delta l,t)\right|}{\chi_a(\Delta l,t)}\mathrm{,}
\end{equation}
where $\Delta l=\Delta x=\Delta y$, $\chi$ represents the mass or the perimeter and the subscript $n$ and $a$ refer to the numerical and analytical solution.} For both observables, the order of convergence is between $1.75$ and $2.0$. Similar results have been obtained for the other density ratios considered here.
\subsection{Isothermal droplet evaporation}\label{subsec:case2}
This test case considers the evaporation of a two-dimensional circular droplet, using the same domain and boundary conditions as in the previous section. The difference is that the evaporation is driven by a difference between the species concentration at the interface and the domain boundary. Hence, we now solve eq.~\eqref{mass_fraction_n} with Dirichlet boundary conditions $Y_{2,\infty}^l = 0$ at the domain boundaries, and $Y_{2,\Gamma}^l = 0.5$ at the interface. The mass flux $\dot{m}$ is then computed from eq.~(\ref{eqn:mass_flux_2}). This allows us to assess the accuracy of the method for coupling the vapor species transport to the Navier-Stokes equations, decoupled from the energy equation. The other physical parameters are reported in table~\ref{tbl:table_groups}, and are chosen to match the initial shrinking rate of the setup presented in the previous section. The steady state solution of eq.~\eqref{mass_fraction} with $\mathbf{u}=0$ is used as initial condition for $Y_2^l$.\par
For this system one can derive an ordinary differential equation for the droplet diameter $d$, assuming a static droplet in a circular domain with diameter $D$ \cite{irfan2017front}:
\begin{equation}
    \frac{\mathrm{d}d^2}{\mathrm{d}t\,\,\,}=-\frac{8\rho_1D_{lg}}{\rho_2}\frac{\ln(1+B_Y)}{\ln(D/d)}\mathrm{,}
	\label{eqn:an_sol}
\end{equation}
where $B_Y=(Y_{2,\Gamma}^l-Y_{2,\infty}^l)/(1-Y_{2,\Gamma}^l)$ is a mass number. In the present setup $D=4\,d_0$ is the domain length, and $B_Y=1$. The equation above can therefore be solved numerically, yielding a reference solution for $d(t)$.\par
\ns{
\begin{figure*}[h!]
	\centering
	\includegraphics[height=4.11 cm,width=\textwidth]{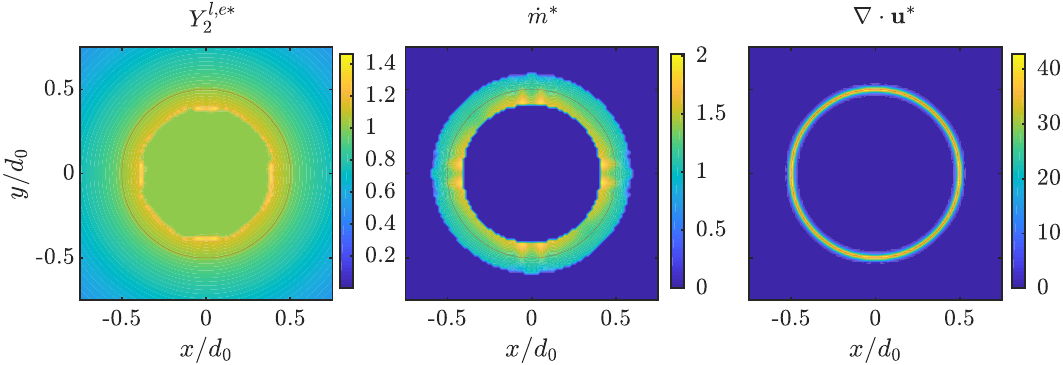}
    \put(-240,0){\small(\textit{a})}
    \put(-125,0){\small(\textit{b})}
    \put( -10,0){\small(\textit{c})}
    \caption{\ns{Isothermal evaporation of a static droplet. (\textit{a}): contours of the normalized extended vapor mass fraction field $Y_{2}^{l,e*}=(Y_{2}^{l,e}-Y_{2,\infty}^{l,e})/(Y_{2,\Gamma}^{l,e}-Y_{2,\infty}^{l,e})$; (\textit{b}): contours of the dimensionless interfacial mass flux $\dot{m}^*=\dot{m}d_0/(\rho_2D_{lg})$ and, (\textit{c}): contours of the dimensionless mass source term $\nabla\cdot\mathbf{u}^*=(d_0^2/D_{lg})\dot{m}(1/\rho_2-1/\rho_1)\delta_\Gamma$ at $tD_{lg}/d_0^2=0.065$. The interface location is depicted by the solid red line. The contour plots \ns{are} restricted to the sub-domain $[-0.75\, d_0,0.75\, d_0]^2$.}}
	\label{fig:3panel_case_2a}
\end{figure*}
Figure~\ref{fig:3panel_case_2a}(\textit{a}) reports the non-dimensional extended vapor mass fraction field for the case of $\lambda_{\rho}=100$ evaluated at the highest resolution ($256\times 256$). In the gas phase, $Y_{2}^{l,e}$ is equal to the vapor mass field, whereas inside it is computed with the procedure explained in section~\ref{num:mass_calc} and only in the region where $0<C<1$. This field is used to compute the interfacial mass flux  depicted in figure~\ref{fig:3panel_case_2a}(\textit{b}), to advect the interface at the next time-step and as a source term in the pressure Poisson equation. As shown in figure~\ref{fig:3panel_case_2a}(\textit{c}),
 this source term is zero everywhere except in a narrow region across the interface. The proposed approach leads to good numerical results in term of mass and droplet shape conservation when compared to the analytical solution, as shown next.}

\begin{figure*}[t!]
    \centering
    \begin{subfigure}[t]{0.5\textwidth}
        \centering
        \includegraphics[width=\textwidth,height=5.2 cm]{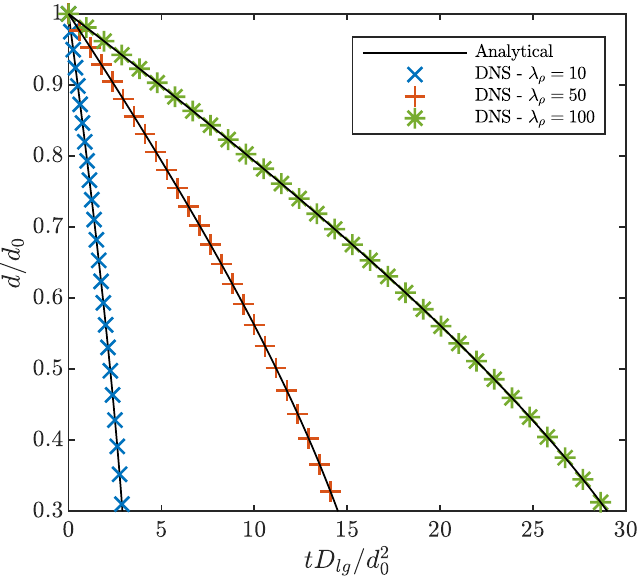}
        %\caption{}
    \end{subfigure}%
    ~ 
    \begin{subfigure}[t]{0.5\textwidth}
        \centering
        \includegraphics[width=\textwidth,height=5.2 cm]{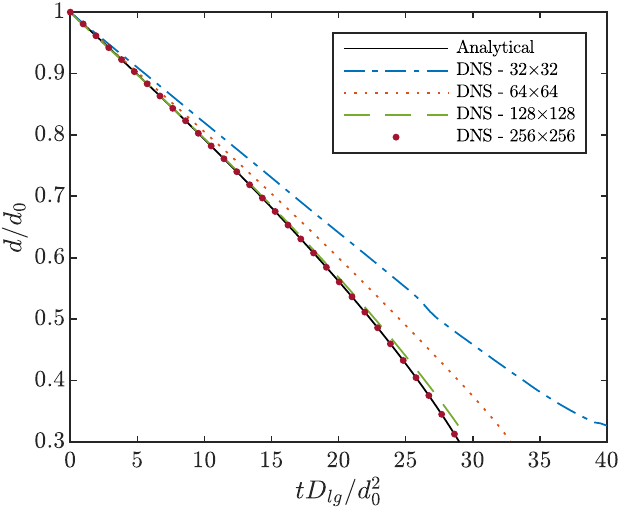}
        %\caption{}
    \end{subfigure}
    \put(-190,0){\small(\textit{a})}
    \put( -10,0){\small(\textit{b})}
    \caption{(\textit{a}): temporal evolution of the normalized droplet diameter for varying $\lambda_{\rho}$, on a $256\times 256$ grid. (\textit{b}): grid convergence test for $\lambda_{\rho}=100$. The DNS discretizes the diffusion term in eq.~\eqref{mass_fraction_n} for $Y_2^l$ implicitly.}%\pc{don't forget to plot $d/d_0$ instead.}}
	\label{fig:iso_t_imp}
\end{figure*}

\begin{figure*}[t!]
    \centering
    \begin{subfigure}[t]{0.5\textwidth}
        \centering
        \includegraphics[width=\textwidth,height=5.2 cm]{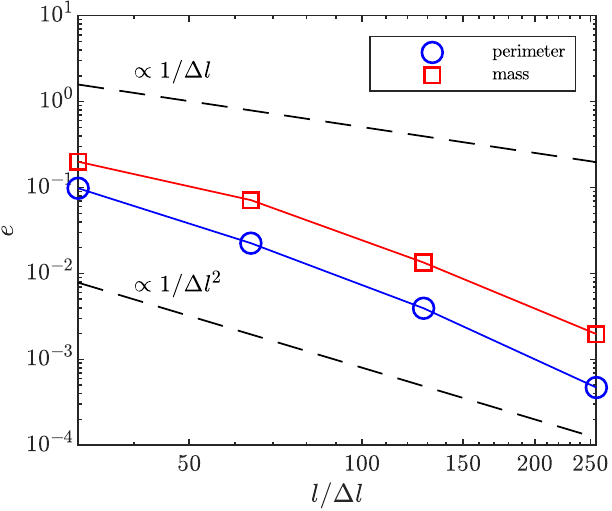}
        %\caption{}
    \end{subfigure}%
    ~ 
    \begin{subfigure}[t]{0.5\textwidth}
        \centering
        \includegraphics[width=\textwidth,height=5.2 cm]{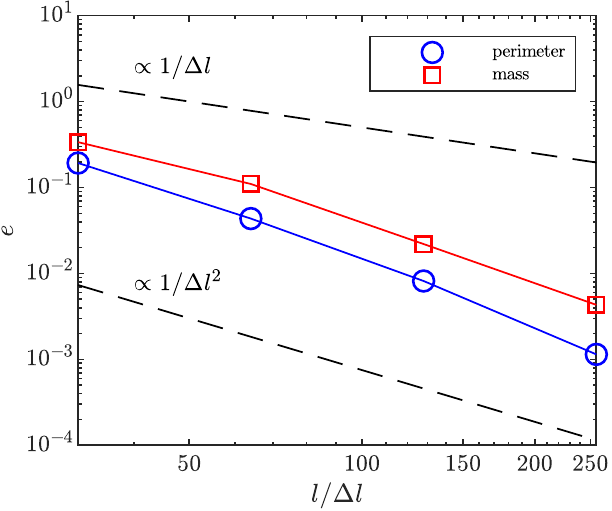}
        %\caption{}
    \end{subfigure}
    \put(-190,0){\small(\textit{a})}
    \put( -12,0){\small(\textit{b})}
    \caption{Error in the normalized droplet mass, $M/M_{t=0}$, and perimeter, $P/P_{t=0}$ for $\lambda_{\rho}=100$ and $d/d_0=0.4$. $\Delta l$ and $l$ denote the grid spacing and domain length. The two panels pertain to simulations with (\textit{a}) implicit and (\textit{b}) explicit treatment of the vapor mass diffusion term in eq.~\eqref{mass_fraction_n}.}
	\label{fig:o_eva_100_case2}
\end{figure*}

Figure~\ref{fig:iso_t_imp}(\textit{a}) presents the time history of the normalized droplet diameter, computed from its volume (i.e., in $2D$, $d=\sqrt{4\,A/\pi}$ where $A$ is the surface area), for different values of $\lambda_\rho$, on the finest grid considered. The solution grid convergence is shown in panel (\textit{b}) of the same figure, for $\lambda_\rho=100$. \ns{Once more, the results show excellent agreement with the reference data, although, as expected, the deviations from the analytical solution are larger than those reported in the previous section as the interfacial mass flux $\dot{m}$ is here computed from the vapor mass gradient and not imposed.} Though these cases correspond to an implicit discretization of the diffusion of $Y_2^l$ in eq.~\eqref{mass_fraction_n}, similar results are obtained with an explicit treatment of the diffusion term. Figure~\ref{fig:o_eva_100_case2} reports the convergence of the numerical error with grid refinement with (panel \textit{a}) and without (panel \textit{b}) implicit diffusion for $\lambda_\rho=100$, $d/d_0=0.4$ and fixed time step $\Delta t=0.005\,d_0^2/D_{lg}$. As we can see, the numerical error is slightly larger when the vapor mass diffusion is discretized explicitly. Yet, both approaches show an error convergence between $1.75$ and $2.0$ as the grid is refined. \par

\subsection{Fully-coupled system -- reproduction of psychrometric data}\label{subsec:case3}
Next we consider the solution of the fully coupled system in the same configuration of the previous sections. In this case, evaporation is driven by the partial pressure of the vaporized liquid near the interface, which is smaller than the corresponding saturated value at the interface $p_{2,\Gamma}^{l,sat}$. We recall that $p_{2,\Gamma}^{l,sat}$ is related to the interface temperature through the Clausius-Clapeyron relation (eq.~\ref{eqn:bc_y}).\par
We consider a stationary two-dimensional circular droplet with the same geometry and outflow conditions of the previous cases. The energy equation is solved with Dirichlet temperature boundary conditions $T_\infty$ at all boundaries. As in the previous case, the steady state solution of eq.~\eqref{mass_fraction} with $\mathbf{u}=0$ is used as initial condition for $Y_2^l$, with Dirichlet boundary conditions $Y_{2,\infty}^l$ at the domain boundaries, and now with prescribed interfacial value $Y_{2,\Gamma}^l$ computed from eqs.~\eqref{eqn:bc_y} and~\eqref{eqn:eq_state}.\par
Similarly to the recent work in \cite{irfan2017front}, we use psychrometric data to validate our numerical method. A water droplet is immersed in air with relative humidity $\psi$ and so-called dry bulb temperature, $T_{db}$, equal to the initial droplet temperature. As evaporation is triggered, the droplet is cooled and its temperature decreases. This evaporative cooling is counterbalanced by a conductive heat flux from the warmer air to the droplet. Eventually, a uniform equilibrium temperature is reached inside the droplet -- so-called wet-bulb temperature, $T_{wb}$. Accordingly, in our computational domain the temperature boundary condition is set to the desired dry bulb temperature $T_\infty = T_{db}$, and the corresponding mass fraction prescribed at the boundary, $Y_{2,\infty}^l$; this is computed from the desired air relative humidity $\psi$ at the dry bulb temperature $T_{db}$ (cf.\ eq.~\ref{eqn:bc_y}): 
\begin{equation}
    Y_{2,\infty}^l=\frac{\psi p_{2,\infty}^{l,sat}M_l}{(p_t-\psi p_{2,\infty}^{l,sat})M_g+\psi p_{2,\infty}^{l,sat}M_l}
\end{equation}
with $p_{2,\infty}^{l,sat}$ computed from eq.~\eqref{eqn:eq_state} evaluated at $T_{db}$.\par
For simplicity, in addition to the non-dimensional governing parameters in table~\ref{tbl:table_groups}, we report the system properties in table~\ref{tbl:thphy_air_water} and its caption.
\ns{As in~\cite{irfan2017front}, the liquid density and thermal conductivity are set smaller than those of water, while keeping the thermal diffusivity constant. The reasons behind this choice are twofold. First, the present methodology becomes less accurate as the density ratios increases (as for most of interface-capturing methods) and using $\lambda_{\rho}=1000$ would require to greatly reduce the time-step. Second, the interfacial mass flux decreases almost linearly when increasing the density ratios and therefore, the time to reach the wet-bulb temperature in this configuration where convection effects are limited would be unnecessarily long to test the method.}\par
\begin{figure*}[t!]
	\centering
	\includegraphics[height=3.6 cm,width=\textwidth]{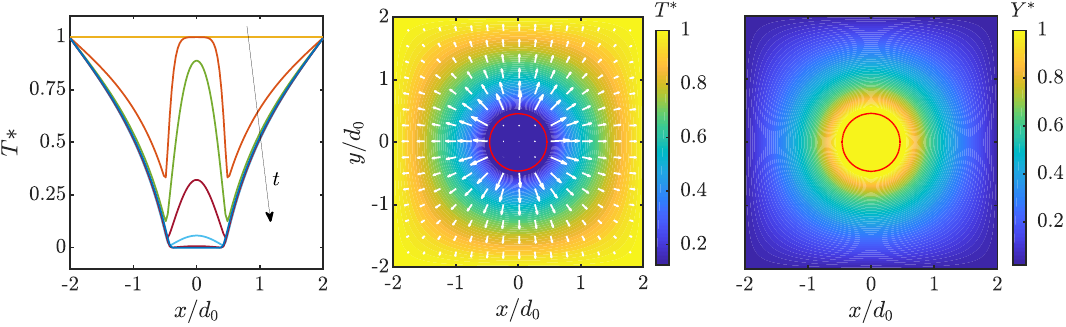}
    \put(-240,0){\small(\textit{a})}
    \put(-125,0){\small(\textit{b})}
    \put( -10,0){\small(\textit{c})}
    \caption{Evaporating static droplet with $T_{db}=313\, \mathrm{K}$ and $\psi=50\%$. (\textit{a}): time evolution of the normalized temperature profile $T^{*}=(T-T_{wb,0})/(T_{\infty}-T_{wb,0})$ along $x=0$ at $tD_{lg}/d_0^2=\lbrace 0,\, 0.95,\, 3.80,\, 11.4,\, 22.8,\, 34.2,\,45.6\rbrace$. (\textit{b,c}): velocity field with contours of $T^*$ (\textit{b}), and contours of normalized vapor mass fraction $Y^{*}=(Y_2^l-Y_{2,\infty}^l)/(Y_{2,\Gamma}^l-Y_{2,\infty}^l)$ (\textit{c}) at $t=45.6\, d_0^2/D_{lg}$.}
    %\textcolor{blue}{non corrisponde??}}%\pc{change axes to -2,2?}}
	\label{fig:tmv_case2b}
\end{figure*}
\begin{figure*}[t!]
    \centering
    \begin{subfigure}[t]{0.5\textwidth}
        \centering
        \includegraphics[width=\textwidth,height=5.2 cm]{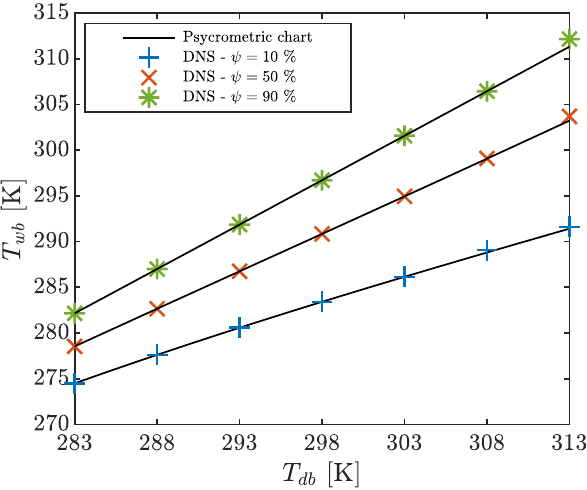}
        %\caption{}
    \end{subfigure}%
    \begin{subfigure}[t]{0.5\textwidth}
        \centering
        \includegraphics[width=\textwidth,height=5.2 cm]{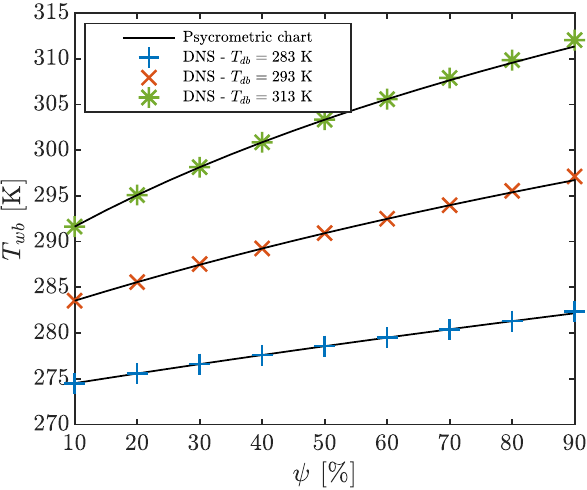}
        %\caption{}
    \end{subfigure}
    \put(-190,0){\small(\textit{a})}
    \put( -12,0){\small(\textit{b})}
    \caption{Numerical results of wet bulb temperature $T_{wb}$ of an evaporating static droplet, compared to the psychrometric chart values varying (\textit{a}) dry bulb temperature $T_{db}$ and  (\textit{b}) relative humidities $\psi$, on a $256\times 256$ grid.}
	\label{fig:psy1}
\end{figure*}

\begin{table}[t!]
\centering
\begin{tabular}{c|ccccc}
   	\toprule
    Phase           & $\rho\,\mathrm{[kg/m^3]}$ & $\mu\,\mathrm{[kg/(m\,s)]}$ & $c_p\,\mathrm{[J/(kg\, K)]}$ & $k\,\mathrm{[W/(m\, K)]}$ & $M\,\mathrm{[kg/kmol]}$ \\
 	  \midrule
   	Gas    & $1.2$  &  $1.79\cdot 10^{-5}$ &  $1006$ & $0.026$ & $29.0$\\
   	Liquid & $10$   &  $1.14\cdot 10^{-3}$ &  $4186$ & $0.006$ & $18.0$\\
    \bottomrule
    \end{tabular}
    \caption{Thermophysical properties of the two-phase system used for comparison with psychrometric data. Other governing parameters are: $d_0=2.5\cdot 10^{-4}$\,$\mathrm{m}$, $T^{sat}=373.15\,\mathrm{K}$, $h_{lv}=2.33\cdot 10^6\, \mathrm{J/kg}$, $\sigma=0.072\,\mathrm{N/m}$ and $D_{lg}=2.23\cdot 10^{-5}\,\mathrm{m^2/s}$.}
\label{tbl:thphy_air_water}
\end{table}
Figure~\ref{fig:tmv_case2b}(\textit{a}) illustrates the time evolution of the droplet temperature profile, for $T_{db}=313\,\mathrm{K}$ and $\psi=50\%$. Indeed, an equilibrium droplet (wet bulb) temperature $T_{wb}$ is attained after a transient due to the mechanisms described above. The velocity vector field, temperature and mass fraction at this equilibrium condition are shown in panels (\textit{b},\textit{c}) of the same figure.
We performed simulations for several combinations of $T_{db}$ and $\psi$, and compared the resulting equilibrium droplet temperature to the expected value of the wet-bulb temperature $T_{wb}$ from psychrometric data, see e.g.\ \cite{herrmann2009thermodynamic}. Figure~\ref{fig:psy1} displays the numerical results of wet bulb temperature to psychrometric data for several combinations of $\psi$ and $T_{db}$, on the finest grid considered ($256\times 256$). The agreement is very good, with slight deviations for the highest temperature and relative humidity considered. A similar trend has been reported in \cite{irfan2017front}, where the discrepancies have been attributed to the assumption of constant thermophysical properties, which actually vary in a not negligible way at high values of $T_{db}$ and $\psi$.\par
\begin{figure*}[t!]
    \centering
    \begin{subfigure}[t]{0.5\textwidth}
        \centering
        \includegraphics[width=\textwidth,height=5.2 cm]{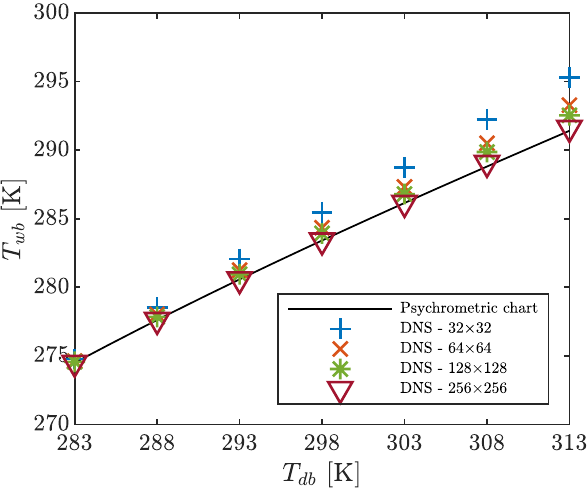}
        %\caption{}
    \end{subfigure}%
    \begin{subfigure}[t]{0.5\textwidth}
        \centering
       \includegraphics[width=\textwidth,height=5.2 cm]{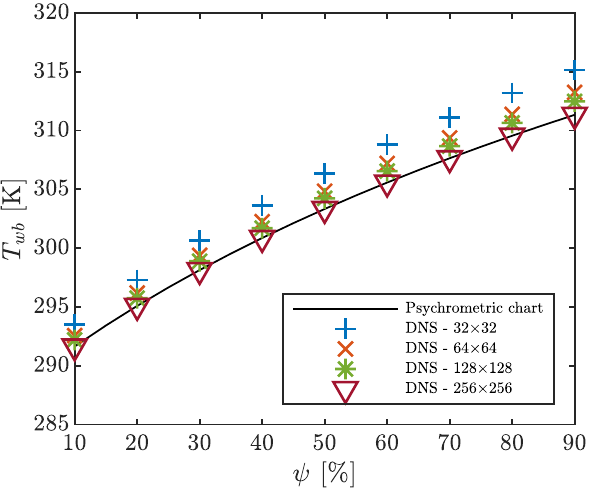}
        %\caption{}
    \end{subfigure}
    \put(-190,0){\small(\textit{a})}
    \put( -12,0){\small(\textit{b})}
    \caption{Grid convergence study for the wet bulb temperature $T_{wb}$ of an evaporating static droplet: (\textit{a}) fixed relative humidity $\psi=10\,\%$, and  (\textit{b}) fixed dry bulb temperature $T_{db}=283\,\mathrm{K}$.}
	\label{grid_conv}
\end{figure*}
In figure~\ref{grid_conv} we present the grid convergence of the solution pertaining the target wet bulb temperature $T_{wb}$ for varying $T_{db}$ (panel \textit{a}) and $\psi$ (panel \textit{b}). Despite the larger grid-sensitivity of the results at higher temperatures for coarser grids, all cases converge towards the expected value. 
\begin{figure*}[t!]
    \centering
    \begin{subfigure}[t]{0.5\textwidth}
        \centering
        \includegraphics[width=\textwidth,height=5.2 cm]{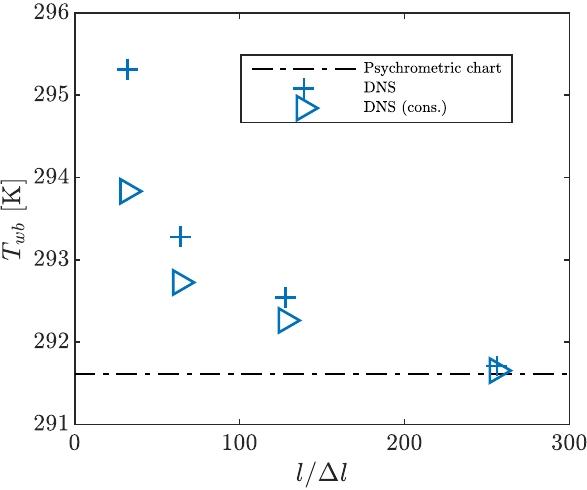}
        %\caption{}
    \end{subfigure}%
    \begin{subfigure}[t]{0.5\textwidth}
        \centering
       \includegraphics[width=\textwidth,height=5.2 cm]{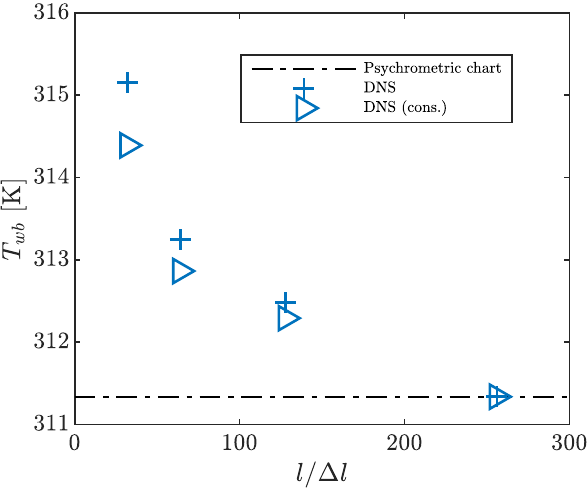}
        %\caption{}
    \end{subfigure}
    \put(-190,0){\small(\textit{a})}
    \put( -12,0){\small(\textit{b})}
    \caption{\ns{Comparison between the non-conservative and the conservative discretization (the latter labeled with \textit{cons.} in the figure) of the energy equation for the numerical calculation of the wet-bulb temperature: (\textit{a}) dry-bulb temperature and relative humidity $T_{db}=313\,\mathrm{K}, \psi=10\,\%$ and, and (\textit{b}) dry-bulb temperature and relative humidity $T_{db}=283\,\mathrm{K}, \psi=90\,\%$.}}
	\label{grid_conv_con}
\end{figure*}

\ns{Finally, we further investigate the large deviations that occur at higher $T_{db}$ and $\psi$ by performing a comparison between the fully conservative discretization and the non-conservative discretization of the energy equation. As figure~\ref{grid_conv_con} shows, when a conservative discretization is employed 
the error is clearly reduced at lower resolution, whereas the difference is less appreciable when a highly refined grid is employed.}

\subsection{Sedimenting droplet in a confining container}\label{subsec:case4}
This configuration illustrates the ability of the method to handle droplet evaporation under pronounced deformations, with close interaction with solid boundaries and large temperature gradients. We consider an evaporating droplet with initial diameter $d_0$ flowing down a confining container with hot walls under the action of gravity, acting in the negative $y$ direction. The droplet is initially at rest in a domain with dimensions $[0,2.5\, d_0] \times [0,10\, d_0]$, in an off-centered position at the top of the container $\mathbf{x}_c=[0.25 ,\, 0.9 ]\, d_0$. No-slip and no-penetration (wall) boundary conditions are prescribed at all boundaries, except for the top side, where a zero-pressure outflow is prescribed. Dirichlet boundary conditions for temperature ($T_w$) and mass fraction ($Y_{2,w}^l=0$) are prescribed at the walls, and (zero) Neumann at the outflow. The initial temperature field is uniform with temperature $T_0=0.95\, T^{sat}<T_w$, whereas the mass fraction field is initialized as described in the previous section. The relevant flow parameters are reported in table~\ref{tbl:table_groups}. Three cases are considered: a droplet with (1) mild and (2) high evaporation rates, and (3) a reference case without phase change. These different evaporation rates are achieved by varying the wall temperature while keeping the other parameters constant, namely $T_w=1.15\, T^{sat}$ and $T_w=2.54\, T^{sat}$ for the two phase-changing cases. To better qualify the evaporation regime, we define a wall Stefan number based on the difference between the imposed wall temperature and the initial temperature, i.e., $\mathrm{Ste}_w\equiv c_{p,2}(T_w-T_0)/h_{lv}$.\par

\begin{figure}[t!]
	\centering
	\includegraphics[height=7.0 cm,width=7.114 cm]{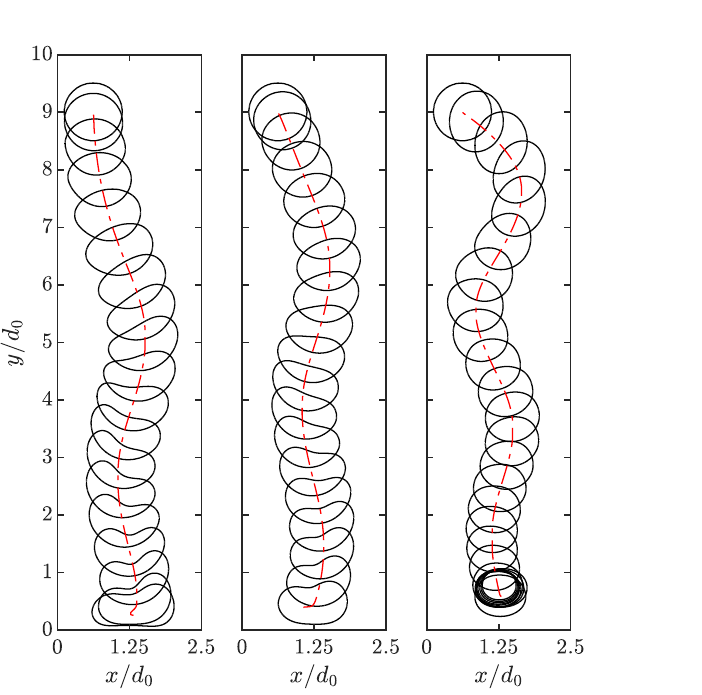}
    \caption{\ns{Trajectory of the droplet's center of mass} (red dashed-dotted line) of a sedimenting droplet in a confining container on a $128\times 512$ grid for (from left to right) $\mathrm{Ste}_w=0$ (i.e.\ no phase change), $\mathrm{Ste}_w=0.033$ and $\mathrm{Ste}_w=0.26$. The interface is shown by the solid black lines with a time interval equal to $0.67\sqrt{d_0/|\mathbf{g}|}$.}
	\label{fig:vof_case3}
\end{figure} 

\begin{figure}[h!]
	\centering
	\includegraphics[height=7.0 cm,width=7.114 cm]{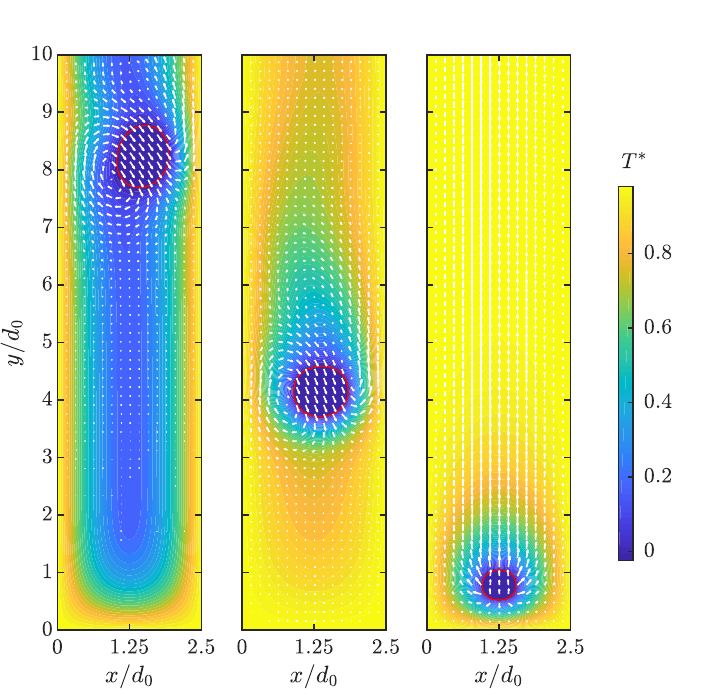}
    \caption{Evaporation of a sedimenting droplet in a confining container at $\mathrm{Ste}_w=0.26$ on a $128\times 512$ grid. Contour plots of normalized temperature ($T^{*}=(T-T_0)/(T_w-T_0)$) at three time-instants (from left to right): $t\sqrt{|\mathbf{g}|/d_0}=1.70$, $6.75$ and $26.9$. The interface location is depicted by the solid red line.}
	\label{fig:tmv_case3}
\end{figure} 

Figure~\ref{fig:vof_case3} depicts the time history of the droplet sedimentation for the three cases under consideration. All cases show an oscillatory trajectory, due to droplet-wall interactions in this confined geometry and to the droplet inertia. As the evaporation rate (i.e.\ $\mathrm{Ste}_w$) increases, the frequency of the droplet oscillation grows, due to the wall-repelling force caused by the Stefan flow. This is particularly evident for the case with $\mathrm{Ste}_w=0.26$, where the droplet starts moving with an almost horizontal velocity, and remains in a levitating state at late times. The droplet dynamics can be better understood by inspecting the temperature and velocity fields. These are shown in figure~\ref{fig:tmv_case3} for the case with $\mathrm{Ste}_w=0.26$ at three different instants. Since $T_w>T_0$, the temperature at the side of the droplet closest to the wall is larger than that on the other side at early times. The Stefan flow  is therefore larger on the warmer side of the droplet, 
which generates a strong wall-repelling force. 
After some time, heat conduction reduces the temperature gradients in the gaseous phase, and the amplitude of the droplet oscillation reduces. 
Finally, at later times, the droplet reaches the bottom side of the container with significant mass loss, and the Stefan flow is sufficiently high to sustain the droplet weight. The droplet therefore remains in a levitating, Leidenfrost-like state in the middle of the container until its complete evaporation.

\begin{figure*}[t!]
    \centering
    \begin{subfigure}[t]{0.5\textwidth}
        \centering
        \includegraphics[width=\textwidth,height=5.2 cm]{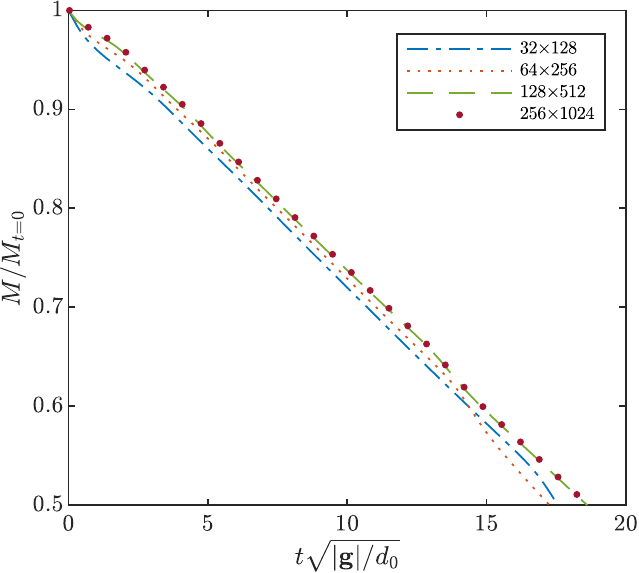}
        %\caption{}
    \end{subfigure}%
    ~ 
    \begin{subfigure}[t]{0.5\textwidth}
        \centering
        \includegraphics[width=\textwidth,height=5.2 cm]{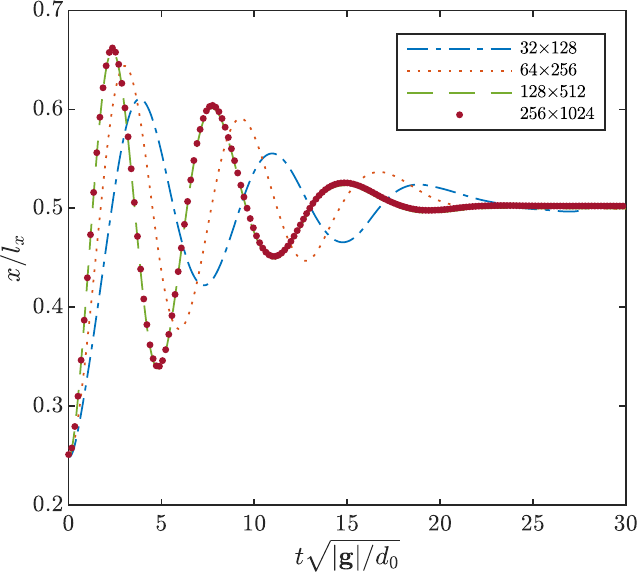}
        %\caption{}
    \end{subfigure}
    \put(-190,0){\small(\textit{a})}
    \put( -12,0){\small(\textit{b})}
    \caption{Grid convergence test for the sedimenting droplet in a confining container with $Ste_w=0.26$, for the droplet mass (\textit{a}), and $x$-component of the trajectory (\textit{b}), showing oscillations along the container centerline.}
	\label{fig:gconv_falling}
\end{figure*}

The solution grid convergence for this more complex case is illustrated for the setup with highest mass transfer, $Ste_w=0.26$, confirming that $128\times 512$ grid points (about $50$ grid points over the initial droplet diameter) suffice for accurately resolving this problem (see figure~\ref{fig:gconv_falling}).

\ns{\subsection{Coalescence of two droplets}\label{subsec:case5}
In this section, we consider the case of two two-dimensional circular droplets interacting and eventually coalescing while evaporating in a hot gas. This test shows the ability of the proposed methodology to handle a complex topological change (i.e., merging), providing stable and convergent results. The two droplets with initial diameter $d_0$ move towards each other with initial velocity $u_{r,0}/2$ in a domain with dimensions $[0,10\,d_0]\times[0,5\,d_0]$, initially, at position 
%$\mathbf{x}_{c,1}$ and $\mathbf{x}_{c,2}$ of the domain, taken equal to 
$[4,\,2.5]d_0$ and $[6,\,2.5]d_0$. 
The relative velocity is chosen to give an impact Reynolds number $Re_{d_0}=\rho_2 u_{r,0}d_0/\mu_2 =25$. The initial temperature field is equal to $T_{g,0}=1.88\, T^{sat}$ in the gas phase, $T_{l,0}=0.95\, T^{sat}$ in the liquid. The mass fraction field is initialized with the steady state solution of eq.~\eqref{mass_fraction}, as done for the previous cases. Zero-pressure outflow boundary conditions are prescribed for velocity and pressure, whereas a Dirichlet boundary condition is prescribed for temperature (equal to $T_{g,0}$) and vapor mass fraction (equal to $Y_{2,\infty}^l=0$). The remaining governing parameters have been reported in table~\ref{tbl:table_groups}, with $C_{\Delta t}=0.1$ for a stable numerical integration.\par
\begin{figure}[t!]
	\centering
	\includegraphics[height=10 cm,width=9 cm]{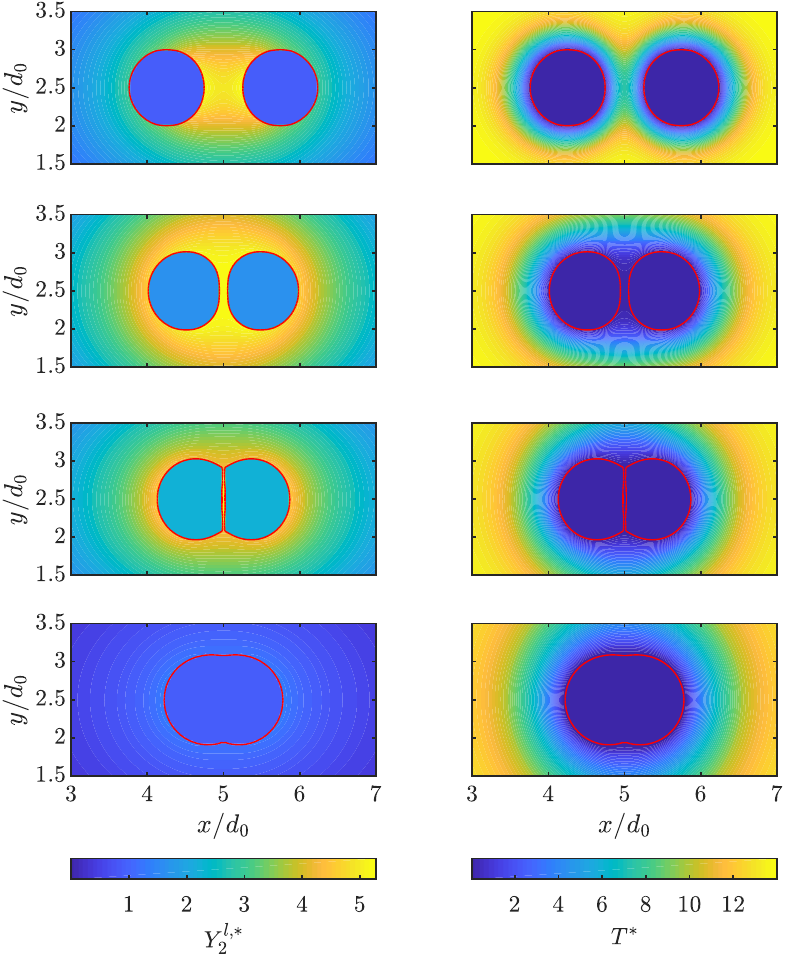}
    \caption{Contours of the vaporised liquid mass fraction $Y_2^{l,*}=(Y_2^{l}-Y_{2,\infty}^l)/(Y_{2,\Gamma}^l-Y_{2,\infty}^l)$ (on the left) and of the dimensionless temperature $T^{*}=(T-T_{\infty})/(T_{wb,f}-T_{\infty})$ at time $tu_{r,0}/d_0=0.78,\, 1.75,\,3.5\,\mathrm{and}\,7.0$ from top to bottom. $T_{wb,f}$ represents the wet bulb temperature of the droplet (uniform from $t^{*}>1$) and $Y_{2,\Gamma}^l$ the interfacial saturation value evaluated at $T_{wb,f}$. The interface is depicted by the iso-contour  $C=0.5$ from the simulation with resolution $512\times 256$. The data are shown in the sub-domain $[3,7]d_0\times [1.5,3.5]d_0$.}
    %\put(-190,0){\small(\textit{a})}
    %\put( -12,0){\small(\textit{b})}
	\label{fig:3p_sca_tmp_case5}
\end{figure} 

\begin{figure}[t!]
	\centering
	\includegraphics[height=5 cm,width=6 cm]{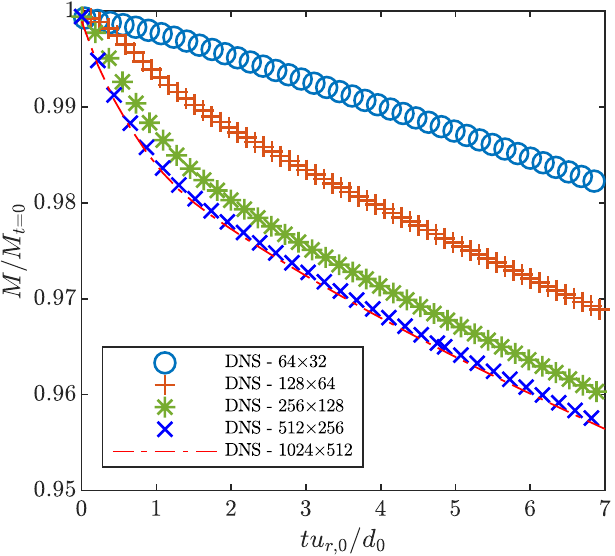}
    \caption{Temporal evolution of the normalized total liquid mass in the system, $M/M_{t=0}$, as a function of the dimensionless time $tu_{r,0}/d_0$ for simulations on different grids (see the figure legend).}
	\label{fig:mass_case_6}
\end{figure} 

Before presenting the results, it is worth recalling that all interface-capturing methods based on the fully Eulerian description of the interface are able to handle automatically topological changes such as coalescence, merging and break-up. This ability comes at the price of having little or no control on the so-called numerical coalescence which occurs when two interfaces are less than one grid cell apart. This is known to over-predict the coalescence rate and can be counteracted by using short-range forces~\cite{de2019effect}, or consider multiple interface fronts as in~\cite{kwakkel2013extension}. Since the influence of the coalescence and merging rate for evaporating droplets is out of the scope of the present work, in the following, we limit to show that the proposed methodology is capable of handling topology changes, providing grid-converging results. \par
The first instances of the simulations are characterized by the highest evaporation rate. As shown in figure~\ref{fig:3p_sca_tmp_case5}, the highest gradients of the mass fraction occur at the early stages when the two droplets are separated and move in opposite direction. Furthermore, the evaporation rate is enhanced by the high temperature gradients between the gas phase and the droplet interface, as it emerges from the time history of the normalized total liquid mass in the system depicted in figure~\ref{fig:mass_case_6}. As the droplets approach and eventually merge, the evaporation rate decreases because the mass transfer of the liquid in the inert gas decreases the temperature around the droplet and thus reduces the saturation value of the mass fraction at the interface. In addition, the droplet velocity decreases and the evaporation process becomes diffusion dominated. \pc{Finally, we note that the simulation is grid convergent, with differences in the history of total liquid mass of less than $1\%$ on the two finest grids ($50$ and $100$ grid points over the droplet diameter).}}

\subsection{Droplet settling in a three-dimensional periodic domain}\label{subsec:case6}
\begin{figure*}[t!]
	\centering
	\includegraphics[width=6.113 cm,height=8 cm]{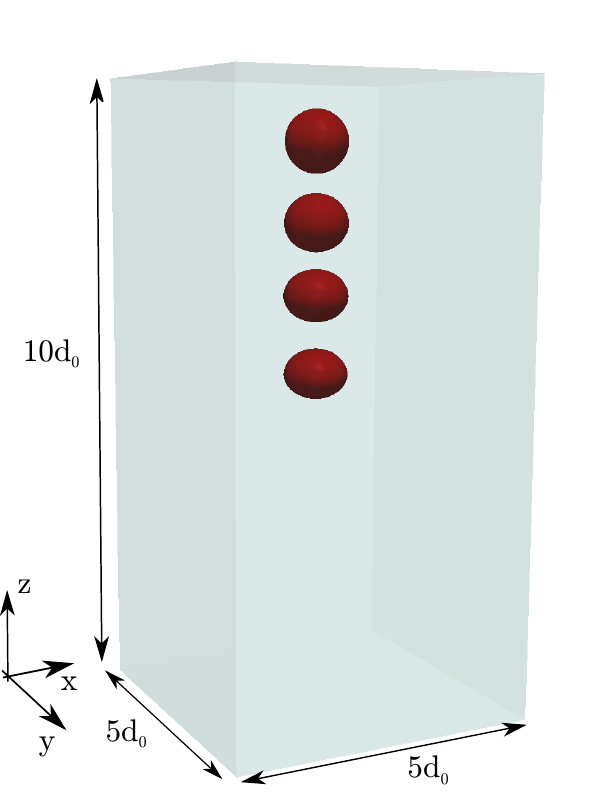}
    \caption{Illustration of the three-dimensional domain with a sedimenting evaporating droplet at four physical times $t\sqrt{|\mathbf{g}|/d_0}=0,\, 0.96,\, 1.91\,\mathrm{and}\, 2.87$. The droplet surface is depicted as the locus of points where $C=0.5$.}
	\label{fig:3d_plane}
\end{figure*}

Finally, we perform a three-dimensional simulation for a fully coupled case, where a single droplet moves, deforms and evaporates in a non-isothermal environment. In this configuration, depicted in figure~\ref{fig:3d_plane}, a spherical droplet with initial diameter $d_0$ settles in a domain with dimensions $[0,5\,d_0]\times[0,5\,d_0]\times [0,10\,d_0]$, under the effect of gravity acting in the negative $z$ direction. The droplet is initially at rest and centered at the top of the domain $\mathbf{x}_c=[2.5, 2.5, 9] d_0$. The domain is periodic in all directions, discretized on a $320\times 320\times 640$ grid. The net weight of the system is subtracted to the fluid momentum balance at each time step to yield zero net acceleration, thereby avoiding a constant acceleration of the entire system \cite{tryggvason2011direct}. The initial temperature field is uniform and given by $T_0=0.95\, T^{sat}$, while the mass fraction field is initialized with the steady state solution of eq.~\eqref{mass_fraction}, as done for the previous cases. The remaining governing parameters have been reported in table~\ref{tbl:table_groups} and we set $C_{\Delta t}=0.1$.
\par

\begin{figure*}[t!]
	\centering
	\includegraphics[width=10.0 cm,height=6.0 cm]{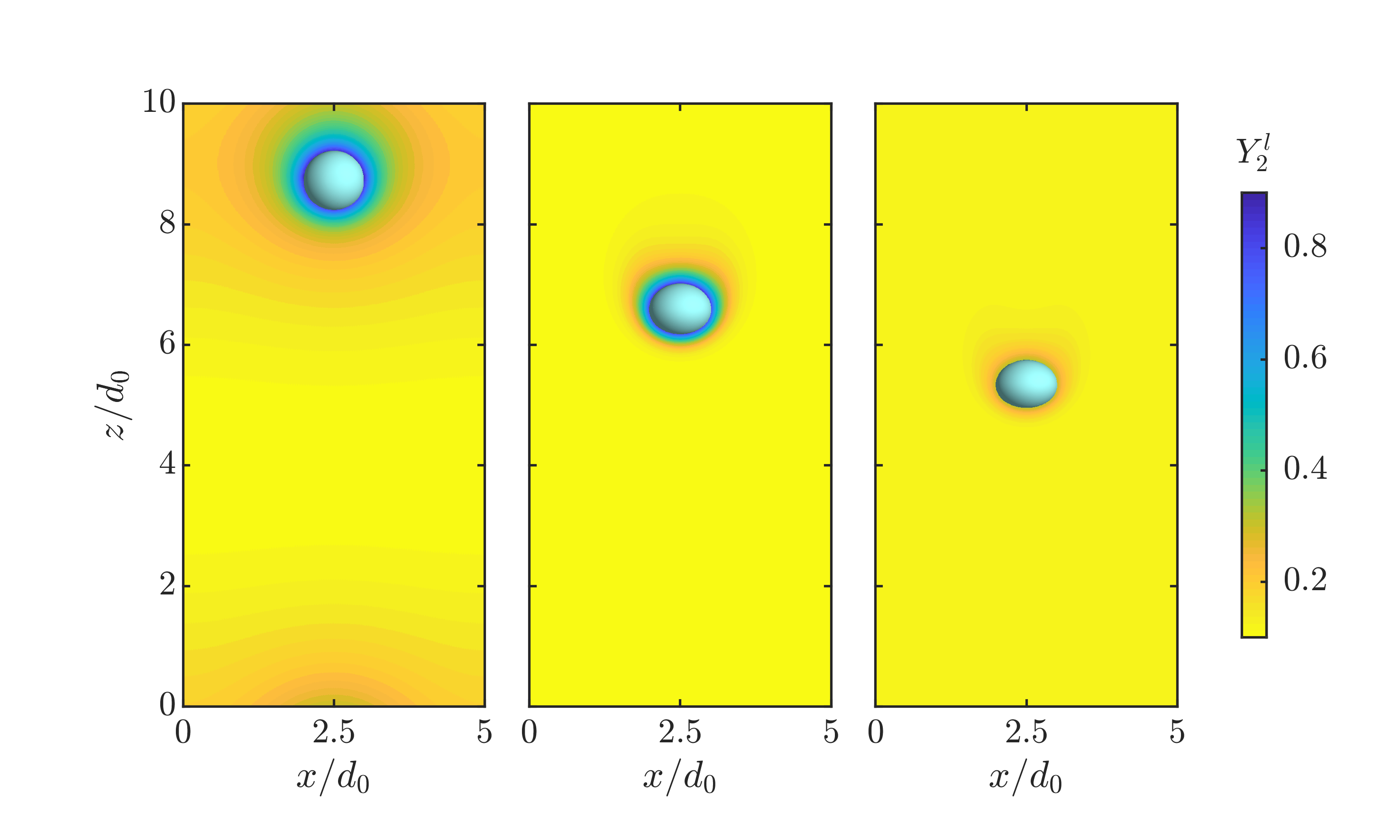}
    \caption{Contours of vaporised liquid mass fraction $Y_2^l$ in the $y=2.5\,d_0$ plane, for time instants $t\sqrt{|\mathbf{g}|/d_0}=0.06,\, 1.91\,\mathrm{and}\,2.87$ (from left to right). The interface is depicted through iso-contours of $C=0.5$.}
	\label{fig:sca_case5}
\end{figure*}

\begin{figure*}[t!]
    \centering
    \begin{subfigure}[t]{0.5\textwidth}
        \centering
        \includegraphics[width=\textwidth,height=5.2 cm]{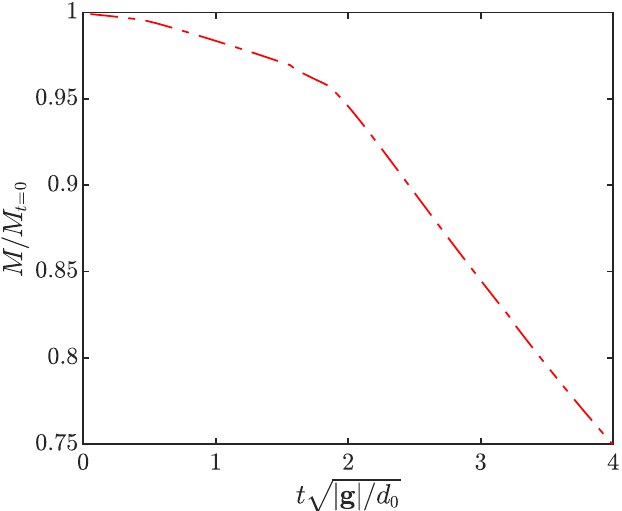}
        %\caption{}
    \end{subfigure}%
    ~ 
    \caption{Temporal evolution of the normalized droplet mass, $M/M_{t=0}$, as a function of the dimensionless time $t\sqrt{|\mathbf{g}|/d_0}$}.
	\label{fig:m_sov_case5}
\end{figure*}

Figure~\ref{fig:sca_case5} presents planar contours of vapor mass fraction for different time instants.
From the two-dimensional distributions of vapor mass fraction in figure~\ref{fig:sca_case5} we qualitatively observe that initially the evaporation process is dominated by diffusion, since the falling velocity is still limited (left panel in figure~\ref{fig:sca_case5}). As time evolves, the droplet accelerates, deforms and starts to lose mass at a faster rate (see the last two panels of figure~\ref{fig:sca_case5}, where the formation of a wake is clear). At this stage, both the increased surface area due to the droplet deformation, and the increasing convective effects tend to increase the mass transfer rate, with the latter effect expected to be more significant. This change in the behavior of the droplet evaporation occurs at $t\sqrt{|\mathbf{g}|/d_0}\approx 2$ as it is clear from the time history of the droplet mass depicted in figure~\ref{fig:m_sov_case5}.

\section{Conclusions}\label{sec:concl}
We have presented a numerical method for interface-resolved simulations of phase changing two-fluid flows using a volume-of-fluid method. The solver is based on an algebraic MTHINC VoF, implemented in an efficient, FFT-based two-fluid finite-difference Navier-Stokes solver. To circumvent the issues related to the jump in fluid velocity across the interface, we transport the VoF function using an interface velocity composed of two parts: (1) a divergence-free extension of the liquid velocity in the entire domain, and (2) a irrotational term accounting for the phase change. This approach requires a simple extension of a standard split advection method with a volume deflation step, and, unlike other approaches in the literature, can be easily applied to other algebraic or geometrical VoF methods. \ns{Furthermore, it can be used in phase-change solvers based on the level-set and conservative level-set method when maintaining the whole domain formulation in solving the Navier-Stokes equations. The divergence-free velocity extension is computed with the aid of a direct fast Poisson solver with negligible computational overhead, about $2\%$ of the total cost.} \par
Evaporation is handled by reconstructing a level-set field from the VoF function, which allows us to benefit from well-established level-set methods for solving phase-change problems; see e.g.\ \cite{tanguy2007level}. The equation of transport of vaporized liquid mass is solved in the gaseous domain, with a Dirichlet boundary condition at the interface, computed from the thermodynamic equilibrium defined by the Clausius-Clapeyron relation. This term can be discretized in time implicitly as in \cite{gibou2002second}, or explicitly. A second-order pde-based extrapolation technique allow us to define the interphase mass flux in a band around the interface, i.e.\ where the VoF function $0<C<1$. As a consequence,  the continuity, momentum and energy equations can be solved with a so-called whole-domain formulation, where source terms accounting for the velocity, stress and heat flux jumps along the interface-normal direction are easily incorporated with a CSF-like approach.\par
The numerical method has been extensively verified and validated against different benchmark cases of increasing complexity. The results illustrate the excellent mass-preserving nature of the method, and the ability of the overall approach to reproduce psychrometric data. Moreover, we show that the method can handle large deformations for a droplet evaporating in the presence of walls, and demonstrate its potential for three-dimensional simulations of evaporating flows. Further, as shown in the \ref{sec:appendixA}, a direct extension of the method can be used to simulate boiling flows. 
\ns{Note finally that the specific implementation presented here might be improved with a discretization of the energy equation in conservative form, and a momentum-preserving method as in \cite{fuster2018momentum} to more efficiently solve problems at high density ratios.}
Overall, we believe that our method has the right ingredients to serve as a base for massive, high-fidelity simulations of phase-changing turbulent flows. 

\section*{Acknowledgements}
The work is supported by INTERFACE, under the project \textit{Hybrid multiscale modelling of transport phenomena for energy efficient processes}, financed by the Swedish Research Council (VR), and by the European Research Council grant, no. ERC-2013-CoG-616186, TRITOS. The computer time was provided by SNIC (Swedish National Infrastructure for Computing) and by the National Infrastructure for High Performance Computing and Data Storage in Norway (project no. NN9561K). St\'ephane Zaleski is acknowledged for the useful discussions, and for pointing out the PhD thesis in \cite{malan2018direct}, which reports a similar approach for constructing a divergence-free velocity extension in the context of boiling flows. 

\appendix

\section{Application of the method for boiling simulations}\label{sec:appendixA}
Here we briefly explain how to extend the numerical framework described for evaporation to study temperature-induced phase change (i.e., boiling), which occurs between a liquid phase and its vapor. From a physical point of view, boiling starts when the partial pressure of liquid in the gaseous phase is equal to the pressure $p_t$ that the surrounding environment exerts on the liquid itself. By fixing $p_t$ and postulating thermodynamic equilibrium at the interface~\cite{ishii2010thermo}, the Clausius-Clapeyron relation indicates that the interfacial temperature $T_{\Gamma}$ is constant and equal to the saturation temperature $T^{sat}$, evaluated at $p_t$. Under this assumption, together with the incompressibility constrain on both phases and weak viscous dissipation, the governing equations reduce to the following form, see e.g.~\cite{ishii2010thermo,juric1998computations},
\begin{subequations}
	\begin{align}
        &\nabla\cdot\mathbf{u}=\dot{m}\left(\dfrac{1}{\rho_2}-\dfrac{1}{\rho_1}\right)\delta_{\Gamma}\mathrm{,}
		\label{eqn:cont_boil}
 \\
        &\rho\left(\dfrac{\partial\mathbf{u}}{\partial t}+\mathbf{u}\cdot\nabla\mathbf{u}\right)=-\nabla p+\nabla\cdot\left(\mu\left(\nabla\mathbf{u}+\nabla\mathbf{u}^T\right)\right)+\rho\mathbf{g}+\sigma\kappa\delta_{\Gamma}\mathbf{n}\mathrm{,}		\label{eqn:mom_boil} \\
%        &\dfrac{\partial(\rho c_p T)}{\partial t}+\nabla\cdot(\rho c_p T\mathbf{u})=\nabla\cdot(k\nabla T)\label{eqn:en_boil}\mathrm{,}
        &\rho c_p\left(\dfrac{\partial T}{\partial t}+\mathbf{u}\cdot\nabla T\right) = \nabla\cdot(k\nabla T)\mathrm{,} \label{eqn:en_boil} \\ % \hspace{2 cm} \textit{with $T_{\Gamma}=T^{sat}}\mathrm{,} \\
        &\dot{m}h_{lv}=k_2\nabla_\Gamma T_2\cdot\mathbf{n}-k_1\nabla_\Gamma T_1\cdot\mathbf{n}\mathrm{;}\label{eqn:mass_boil}
	\end{align}
\end{subequations}
%where eq.~\eqref{eqn:mass_boil} describes the jump in heat flux across the interface due to the latent heat of phase change. 
First, note that the continuity~\eqref{eqn:cont_boil} and momentum~\eqref{eqn:mom_boil} equations are independent of the phase change mechanism. Therefore, their numerical solution follows the procedure reported in section~\ref{num:flow_solver}. Moreover, employing a VoF method to capture the interface dynamics, the interface representation (section~\ref{num:int_repre}) and the interface velocity construction (section~\ref{num:int_vel}) remain formally unchanged, with the minor modification of constructing the interface velocity from a divergence-free extension of the vapor velocity, instead of that of the liquid.\par
To solve the boiling problem, the main difference is that the energy eq.~\eqref{eqn:en_boil} is solved with a Dirichlet boundary condition at the interface: $T_{\Gamma}=T^{sat}$. In practice, we solve eq~\eqref{eqn:en_boil} with the same schemes used for $Y_2^l$, but for a temperature field $T_1$ in $\Omega_1$, and $T_2$ in $\Omega_2$, separately, with $T_{\Gamma}=T^{sat}$ imposed at the interface. \nf{The energy equation is therefore solved in each subdomain with constant termophysical properties (i.e.\ $\rho=\rho_i$, $c_p=c_{pi}$ and $k=k_i$ in $\Omega_i$ ($i=1,2$)}.
Then, following the procedure reported in~\cite{tanguy2014benchmarks}, the temperature field in the liquid domain $T_1$ is extrapolated into the vapor domain $\Omega_2$, and the temperature field in the vapor domain $T_2$ is extrapolated into the liquid domain $\Omega_1$ using the procedure described in section~\ref{num:mass_calc}. The resulting extended fields $T_{1}^{e}$ and $T_{2}^e$ are continuously differentiable across $\Gamma$. Accordingly, they can be used to compute $\dot{m}$ in eq.~\eqref{eqn:mass_boil} using a central difference scheme. This allows to define $\dot{m}$ in a band around the interface, covering the region where the VoF function is $0<C<1$.
%\section*{References}

%\bibliographystyle{plainnat}
\bibliography{bibfile}

\end{document}